\documentclass[draft]{agujournal2019}
\usepackage{url} 
\usepackage{lineno}
\usepackage{soul}
\usepackage{lscape}
\usepackage{longtable}
\usepackage{caption}

\journalname{SMILE Special Issue - Earth and Planetary Physics}

\begin{document}

%
%


\title{Solar wind ion charge state distributions and compound cross sections for solar wind charge exchange X-ray emission}

%
%




\authors{Dimitra Koutroumpa}


\affiliation{1}{LATMOS-OVSQ, CNRS, UVSQ Paris-Saclay, Sorbonne Université, 11 Boulevard d'Alembert, 78280, Guyancourt, France}




\correspondingauthor{Dimitra Koutroumpa}{dimitra.koutroumpa@latmos.ipsl.fr}




\begin{keypoints}
\item Individual ion abundances show strong variations per solar wind type and solar activity period.
\item Calculations with broad energy band cross-sections smooth out individual variations for different solar wind types.
\item Separate ion cross-sections and continuous abundance monitoring are crucial for accurate SWCX modeling and future high resolution analyses.
\end{keypoints}

%
%

%
%


\begin{abstract}
Solar Wind Charge eXchange X-ray (SWCX) emission in the heliosphere and Earth's exosphere is a hard to avoid signal in soft X-ray observations of astrophysical targets. On the other hand, the X-ray imaging possibilities offered by the SWCX process has led to an increasing number of future dedicated space missions for investigating the solar wind-terrestrial interactions and magnetospheric interfaces. In both cases, accurate modelling of the SWCX emission is key to correctly interpret its signal, and remove it from observations, when needed. In this paper, we compile solar wind abundance measurements from ACE for different solar wind types, and atomic data from literature, including charge exchange cross-sections and emission probabilities, used for calculating the compound cross-section $\alpha$ for the SWCX X-ray emission. We calculate $\alpha$ values for charge-exchange with H and He, relevant to soft X-ray energy bands (0.1 - 2.0 keV) for various solar wind types and solar cycle conditions.
\end{abstract}


%
%

%


%
%
%
%

\section{Introduction}

Discovered more than 25 years ago following comet Hyakutake's X-ray observations with the Röntgen Satellite-ROSAT \cite{Lisse1996}, solar wind charge exchange X-ray (SWCX) emission is a relatively new discovery in astrophysics. \citeA{Cravens1997} interpreted the emission as the de-excitation of highly charged solar wind ions that capture electrons from the cometary neutrals. It is now established that the emission is omnipresent in the solar system, where the solar wind interacts with planetary environments, including Mars and Venus \cite{Dennerl2002, Dennerl2002a}, the Earth \cite{Cravens2001}, Jupiter \cite{Cravens1995, Branduardi2007}, Pluto \cite{Lisse2017}, and interstellar neutrals flowing through the heliosphere \cite{Lallement2004}. 

SWCX emission in the Earth's magnetosphere was first acknowledged as a time variable background measured during the ROSAT All-Sky Survey \cite{Snowden1994}, and soon the correlation between the solar wind and SWCX emission was established \cite{Dennerl1997,Freyberg1998,Cox1998, Cravens2001}. In the context of astrophysical studies with ROSAT and subsequent X-ray observatories (e.g. XMM-Newton), the SWCX foreground from the geocorona and the heliosphere is a hindrance to studies of extended astrophysical sources \cite<e.g.,>[]{Kuntz2018}. However, heliophysicists recognized in this mechanism a powerful tool for the global study of the solar wind - planet interactions \cite<see>[for a review]{Sibeck2018}. Indeed, these emissions are proportional to the solar wind ion flux and to the density of the neutral targets. The signal is therefore sensitive to variations in these quantities. In regions of solar wind plasma pileup and/or increased neutral density, such as the subsolar magnetosheath and polar cusps, the emission is enhanced, paving the way for imaging of these key regions of the Sun-Earth system \cite{Robertson2003}. Several space missions currently in development will exploit SWCX imaging of plasma density structures to investigate the coupling between the solar wind and the Earth's magnetosphere, such as ESA's Solar wind Magnetosphere Ionosphere Link Explorer-SMILE mission \cite{Branduardi-Raymont2018}, and NASA's Lunar Environment heliospheric X-ray Imager \cite<LEXI;>[]{Walsh2020}. 

The SWCX emission mechanism is expressed by the following reaction:
\begin{linenomath*}
\begin{equation}\label{eqSWCX}
M + X^{q+} \rightarrow M^+ + X^{(q-1)+*} \rightarrow M^+ + X^{(q-1)+} + \gamma_j
\end{equation}
\end{linenomath*}
where the solar wind source ion $X^{q+}$ captures an electron from the target neutral $M$. This produces a new ion in an excited state $X^{(q-1)+*}$, that de-excites by emitting an X-ray photon $\gamma_j$.

\begin{figure}
\noindent\includegraphics[width=\textwidth]{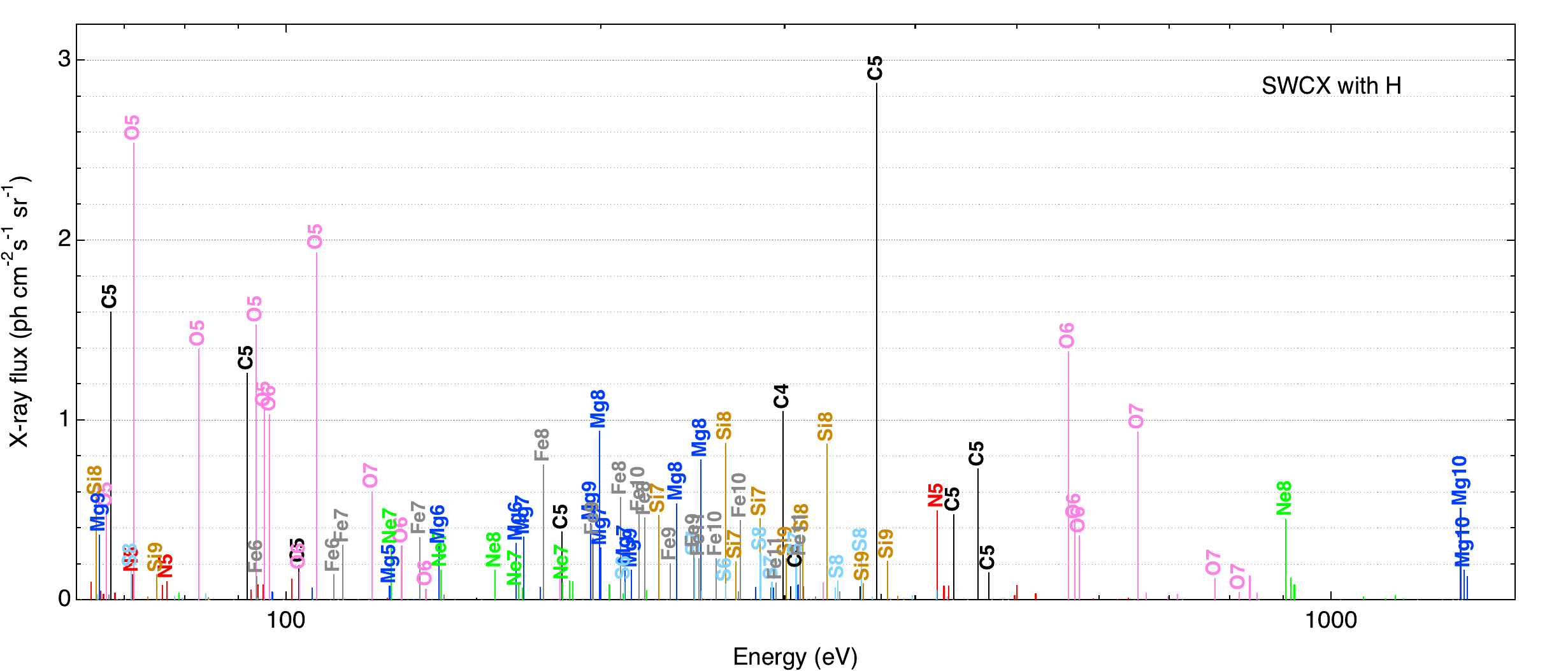}
\noindent\includegraphics[width=\textwidth]{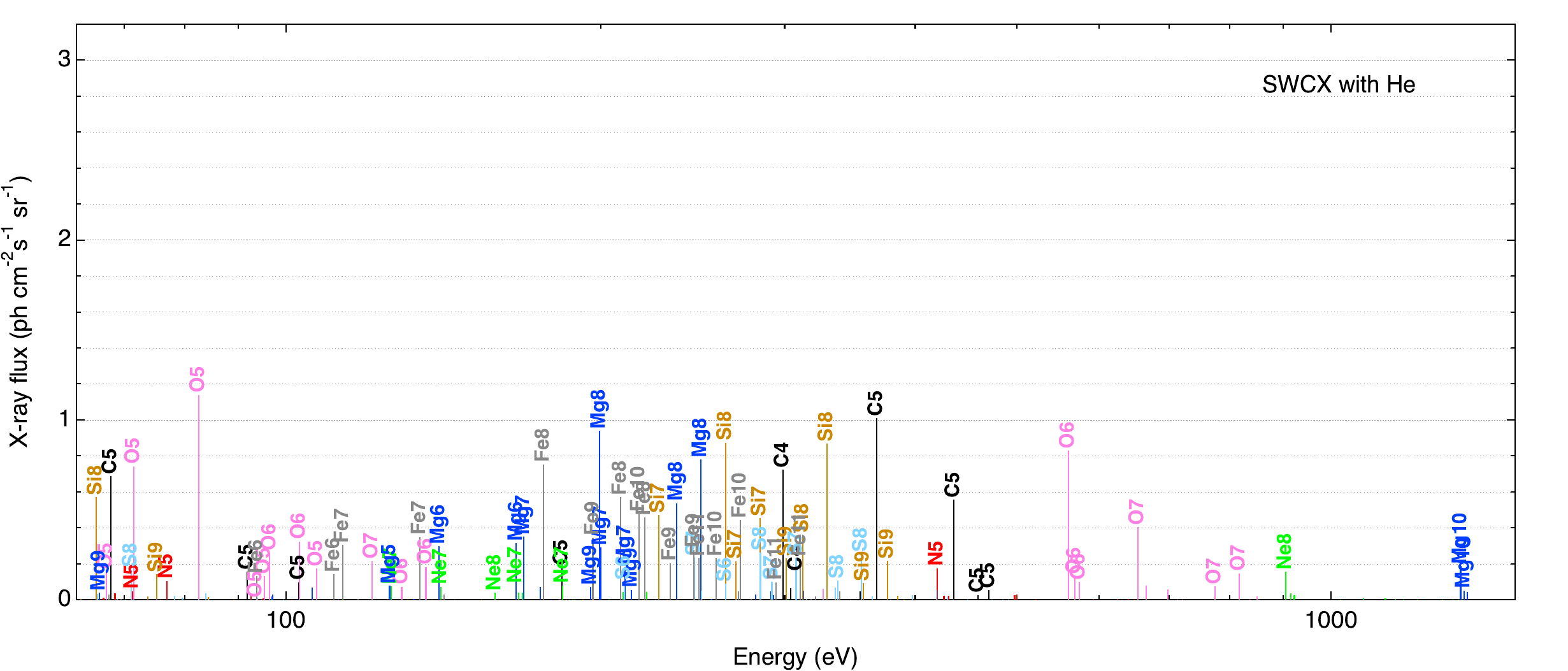}
\caption{Theoretical X-ray spectra produced by SWCX with H (top) and He (bottom), assuming an equal column density for the two neutral targets and typical slow solar wind composition. The emitting ions X$^{(q-1)+}$ (where $q$ refers to the ion charge state) are noted above their corresponding lines. Based on the model by \protect\citeA{Koutroumpa2009b}.}
\label{figSpectra}
\end{figure}

The spectrum of SWCX emission is comprised of discreet spectral lines characteristic of the produced ions $X^{(q-1)+}$ \cite<Figure \ref{figSpectra}, based on>[]{Koutroumpa2009b}. The X-ray flux in a given spectral line is calculated as an integral along the line-of-sight $s$, in units of $photons~ cm^{-2}~ s^{-1}~ sr^{-1}$:
\begin{linenomath*}
\begin{equation}\label{eqXray}
I(\gamma_j) = \frac{1}{4\pi} \int_{s=0}^{\infty} N_M(s)~N_{X^{q+}}(s)~V(s)~\sigma_{X^{q+},M}(V)~Y_{X^{(q-1)+},j}(V)~ds
\end{equation}
\end{linenomath*}
where, $N_M(s)$ is the neutral density, $N_{X^{q+}}(s)$ is the source ion density, $V(s)$ is the ion-neutral collision relative velocity, $\sigma_{X^{q+},M}(V)$ the velocity- and species-dependent cross-section of the collision, and $Y_{X^{(q-1)+},j}(V)$ is the photon emission probability for spectral line $j$ of the produced ion $X^{(q-1)+}$, also dependent on the velocity and neutral target species \cite<Figure \ref{figSpectra}, and>[]{Kharchenko2005}. 

The solar wind ion density is usually expressed as a function of proton density such that $N_{X^{q+}}(s) = \left[\frac{X^{q+}}{p}\right]~N_p(s)$, where $\left[ \frac{X^{q+}}{p}\right]$ is the source ion's abundance relative to solar wind protons. In that case we may assume that the line flux is proportional to the solar wind proton flux $N_p(s)~V(s)$, according to equation \ref{eqXray}. It is generally admitted that the SWCX signal variability is correlated with the solar wind proton flux, especially for broad energy band measurements in the 0.1-0.3 keV energy range (Figure \ref{figSpectra}), where the spectral lines are produced by a multitude of different solar wind ions \cite{Kuntz2015}. However, this is not systematically the case when studying spectral bands dominated by only a few ion species, such as oxygen (0.5-0.7 keV), as demonstrated by \citeA{Kuntz2015}. In several cases, SWCX enhancements were found to be sensitive to increases of ion charge-state abundances rather than, or in addition, to the overall solar wind proton flux enhancement \cite{Snowden2004,Carter2009,Ishi2019,Zhang2022}.

For broad soft X-ray energy band studies with low to moderate spectral resolution, it is convenient to provide a parameter that encompasses the atomic physics parameters (cross sections and emission probabilities) relative to every ion in the solar wind, as well as the composition properties of the later for the specific energy bands. This parameter, called $\alpha$, is defined  in units of $eV~ cm^2$ for SWCX with neutral target $M$ as follows:
\begin{linenomath*}
\begin{equation}\label{eqAlpha}
\alpha_M (V) =  \sum_{X^{q+}}\sum_{\Delta E, j} \left[ \frac{X^{q+}}{p}\right] \sigma_{X^{q+},M} (V) Y_{X^{(q-1)+},j}(V) E_j
\end{equation}
\end{linenomath*}
where $\Delta E$ is the energy range including all spectral lines $\gamma_j$ of energy $E_j$ in units of eV (for example the SMILE/SXI band between 0.1 and 2 keV). In that case, the total X-ray energy flux in the given energy range $\Delta E$ will be the sum of all lines calculated based on equation \ref{eqXray}, such that:
\begin{linenomath*}
\begin{equation}\label{eqband}
R(\Delta E) = \frac{1}{4\pi} \alpha_M \int_{s=0}^{\infty} N_M(s)~N_p(s)~V(s)~~ds
\end{equation}
\end{linenomath*}

The $\alpha$ compound cross-section has been routinely used in SWCX studies, with first empirical estimates varying from 6$\times$10$^{-17}$ to 6$\times$10$^{-15}$ $eV~ cm^2$ for photon energies above 100 eV \cite{Cravens1997,Cravens2000a,Cravens2001}. 

\citeA{Schwadron2000} (hereafter SC00) attempted the first detailed spectroscopic modeling of cometary SWCX emission using a detailed list of solar wind ion charge state abundances from Ulysses data separated into slow and fast wind velocities, and spectral information (rough line energies and cross-sections) based on the approximation that all ions were hydrogen-like \cite{Wegmann1998}. More detailed calculations for the heliospheric SWCX spectrum based on the SC00 abundances were produced by \citeA{Pepino2004} and subsequent studies \cite{Koutroumpa2006,Koutroumpa2009b}. 

The Sun-Earth interaction models fall into several categories, and analysing their specific differences would be beyond the scope of this paper. Magnetohydrodynamic (MHD) codes, such as the Open Geospace Global Circulation Model \cite<OpenGGCM;>[]{Raeder2001}, or the Piecewise Parabolic Method with a LagRangian remap \cite<PPMLR;>[]{Hu2007}, use a fluid description for all plasma components, protons (ions) and electrons alike. These models have the advantage of providing quick computation times and increased spatial and temporal resolution to analyse the plasma dynamics in the Earth's magnetosphere. However, they cannot grasp the kinetic effects that the particles' gyro-motion produce in the presence of magnetic fields. Hybrid \cite<e.g., Latmos Hybrid Simulation-LatHyS;>[]{Modolo2016} and test-particle models \cite{Tkachenko2021} use an approach where protons/ions are described kinetically, allowing for a better description of the kinetic effects. However, these models are more computationally intensive, and counter this drawback with reduced spatial and/or temporal resolution, and/or reduced simulation domains. 

MHD codes produce proton fluxes, then base the SWCX calculations on the proportionality to this quantity as well as the compound cross-section $\alpha$ as shown in equation \ref{eqband} \cite{Sun2019,Connor2021}. On the other hand, test-particle models \cite{Tkachenko2021} have the advantage to calculate the SWCX emission for every ion species individually, while being more time consuming, and less flexible in terms of temporal variability. Both approaches will be complementary to support the science return of the SMILE/SXI data. The former will allow a detailed dynamic study of the general variability of the SWCX signal, while the later approach will allow more precise spectral studies of the SWCX emission, and the effects that the dynamics of individual ions may have on the morphology of the emission around magnetospheric boundaries. 

In this paper we aim to provide solar wind composition estimates for various solar wind conditions as a reference guide to SWCX spectral models, and calculate the compound cross-sections $\alpha$ for various energy bands to assist magnetospheric SWCX simulations. 
In section \ref{SecSW} we present updated solar wind ion composition data from the Advanced Composition Explorer \cite<ACE;>[]{Gloeckler1998} and classify the different solar wind types according to literature, as an extension to the SC00 list. In section \ref{SecAD} we describe the atomic data, including velocity-dependent cross-sections and emission line probabilities. In section \ref{SecResults} we present the results of $\alpha$ for SWCX with H and He atoms in different energy bands and for the various solar wind types, and offer some conclusions in section \ref{SecConclusion}.

\section{Solar Wind Ion composition}\label{SecSW}
The ACE satellite has been monitoring, among other quantities, the solar wind density, velocity and composition from the Lagrange L1 point since 1998. 

\begin{figure}
\noindent\includegraphics[width=\textwidth]{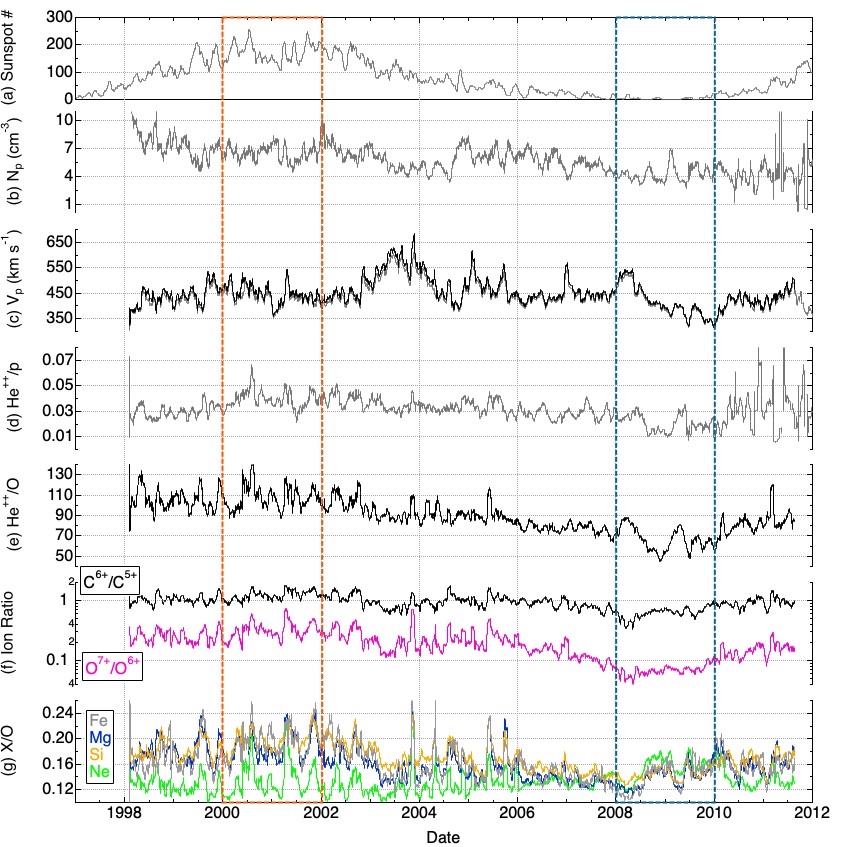}
\caption{From top to bottom: (a) Sunspot number, (b) Proton density from SWEPAM, (c) Proton (grey) and alpha particle (black) velocities from SWEPAM and SWICS respectively, (d) alpha to proton ratio from SWEPAM, (e) alpha to oxygen ratio from SWICS, (f) Carbon (black) and Oxygen (magenta) charge state ratios, (g) Elemental abundances of the heavy ions measured with SWICS (Neon - green, Magnesium - blue, Silicon - yellow, Iron - grey). All quantities are 27-day averages. The orange and blue dashed rectangles mark out respectively the solar maximum and solar minimum limits applied in calculations (See text for details).}
\label{fig1}
\end{figure}

We use the ACE/SWICS 1.1 Level 2 database\footnote{\url{http://www.srl.caltech.edu/ACE/ASC/level2/lvl2DATA_SWICS_SWIMS.html}} from which we extract the ion charge state distributions $[\frac{X^{q+}}{X}]$ and elemental abundances relevant to oxygen $[\frac{X}{O}]$, in order to calculate the ion charge state relative abundances $[\frac{X^{q+}}{O}]$, as well as the alpha ($He^{++}$) particle speed which is equatable with the proton speed. An anomaly that occurred in August 2011 has impacted the ACE/SWICS operations, and detailed charge-state distributions and most elemental abundances are no longer provided since that date. We use the ACE/SWEPAM Level 2 database\footnote{\url{https://izw1.caltech.edu/ACE/ASC/level2/lvl2DATA_SWEPAM.html}} to obtain the proton parameters (density and velocity) and alpha to proton ($\frac{He^{++}}{p}$) ratio. In Figure \ref{fig1} we present a number of parameters showcasing the changes in solar wind properties during the solar cycle, between 1997 and 2012.

\begin{figure}
\noindent\includegraphics[width=0.5\textwidth]{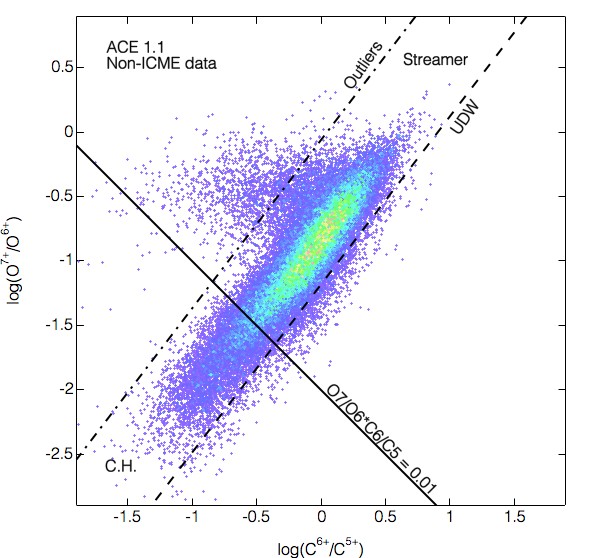}
\noindent\includegraphics[width=0.5\textwidth]{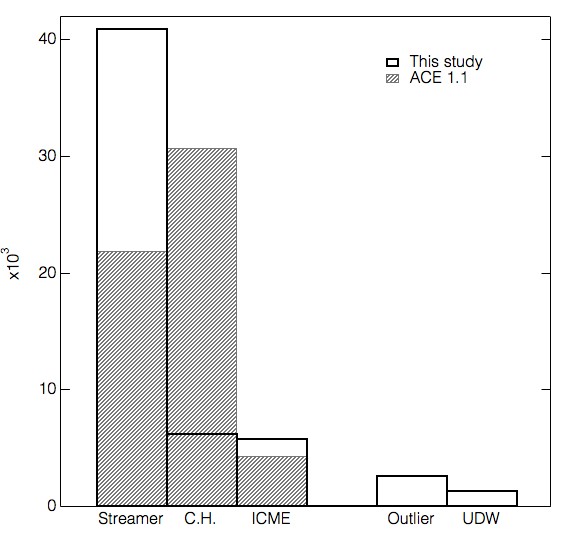}
\caption{Left - 2D histogram of the log($\frac{O^{7+}}{O^{6+}}$) versus the log($\frac{C^{6+}}{C^{5+}}$) ratio from the non-ICME (Interplanetary Coronal Mass Ejection) ACE SWICS 1.1 data. The black lines represent the coronal hole - streamer (solid), the outlier (dot-dashed) and the Upper Depleted Wind - UDW (dashed) type separation (see text for details). Right - Comparison of the solar wind type histogram for this analysis and the ACE 1.1 website classification.}
\label{fig2}
\end{figure}

The ACE 1.1 database provides a parameter that allows a rough classification of the solar wind type, in streamer, coronal hole (CH) and interplanetary coronal mass ejection (ICME), based on the $\frac{O^{7+}}{O^{6+}}$ ratio versus proton speed functions described by \citeA{Zhao2009}\footnote{\url{http://www.srl.caltech.edu/ACE/ASC/level2/ssv4_l2desc.html}}. However, according to \citeA{VonSteiger2015} the $\frac{O^{7+}}{O^{6+}}$ ratio alone, and in particular the threshold ($\frac{O^{7+}}{O^{6+}}\leq0.145$) employed by \citeA{Zhao2009}, is not appropriate to properly differentiate the streamer from CH types. Indeed, upon a closer inspection, the \citeA{Zhao2009} threshold produces an abnormally large population of CH SW type with respect to the other types, which seems unrealistic at ACE's low latitudes \cite<see Figure 1 from>[]{Zhao2009}. \citeA{VonSteiger2015} have demonstrated from Ulysses data that a better parameter to separate streamer from CH wind is the $\frac{O^{7+}}{O^{6+}} * \frac{C^{6+}}{C^{5+}}$ product. A threshold of $\frac{O^{7+}}{O^{6+}} * \frac{C^{6+}}{C^{5+}}\leq0.01$, clearly identifies the CH population from the slow (streamer) solar wind (see Figure 1 from that paper). The ACE data do not show a clear bi-modal distribution as the Ulysses data \cite<compare Figure \ref{fig2}-left to Figure 1 of>[]{VonSteiger2015}, presumably because the Ulysses CH population originates from higher heliolatitudes, as opposed to ACE data measured at low latitudes. This seems to agree with the analysis of \citeA{Zhang2003}, who showed that the $\frac{O^{7+}}{O^{6+}}$ of equatorial CHs seems to have a much broader range of values, compared to the polar CHs that show less scatter (see their figure 5). Even though their intrinsic coronal properties are not significantly different, equatorial CHs are less frequent, short-lived, much smaller in size from polar CHs and their flow speed is lower probably due to deceleration processes from interaction with streamer wind flows. In our case, the $\frac{O^{7+}}{O^{6+}} * \frac{C^{6+}}{C^{5+}}$ product seems to produce a more reasonable distribution of the different SW origins for low latitudes, as demonstrated by the number of CH vs streamer occurrences presented in the histograms in Figure \ref{fig2}-right. 

\citeA{Zhao2017a, Zhao2022} have further investigated the slow solar wind, and have found two more populations that exhibit anomalous composition. The "outlier" solar wind \cite{Zhao2017a} has lower abundances for the bare ions, in particular $C^{6+}$ (Figure \ref{fig2}) and is probably a signature of magnetic reconnection in its source region. The "upper depleted wind" \cite<UDW>[]{Zhao2022}, exhibits systematically depleted elemental abundances, and is most likely associated with quiet Sun regions, while the normal slow wind originates from active regions and the heliospheric current sheet streamers. 

In this analysis we adopt the following classification for the solar wind type: 
\begin{enumerate}
    \item we identify ICMEs based on the list provided by \citeA{Richardson2004}\footnote{\url{https://izw1.caltech.edu/ACE/ASC/DATA/level3/icmetable2.htm}}
    \item we adopt the $\frac{O^{7+}}{O^{6+}} * \frac{C^{6+}}{C^{5+}} \leq 0.01$ threshold to identify the CH (fast) wind, and
    \item we exclude the outlier and UDW slow wind populations from the streamer type.
    
\end{enumerate}
The thresholds applied for the different populations are illustrated in Figure \ref{fig2}-left of the non-ICME wind population.

\begin{figure}
\noindent\includegraphics[width=0.5\textwidth]{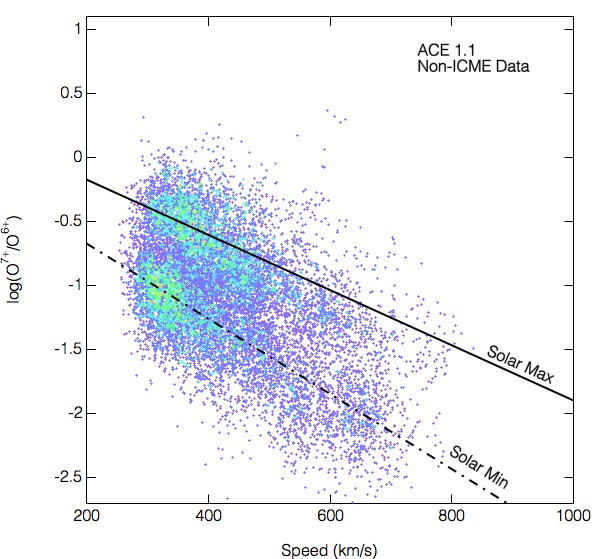}
\noindent\includegraphics[width=0.5\textwidth]{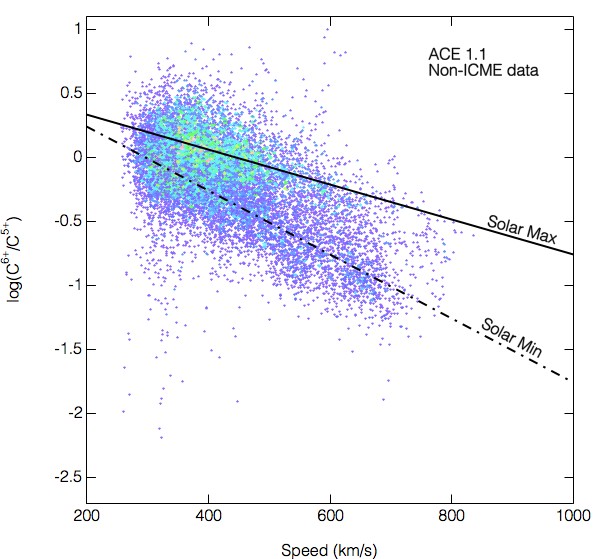}
\caption{Left - 2D histogram of the log($\frac{O^{7+}}{O^{6+}}$)  ratio as a function of velocity for solar maximum (2000-2002) and solar minimum (2008-2010) periods. The linear correlations for solar maximum (solid) and minimum (dot-dashed) are also overplotted, showcasing the sharp difference between the two periods. Right - Same as in the left panel, only for the log($\frac{C^{6+}}{C^{5+}}$) ratio versus the velocity.}
\label{fig3}
\end{figure}

It is also worth noting that ion ratios and elemental abundances in the ACE database show a significant change between solar maximum and solar minimum (see for example panels f, g of Figure \ref{fig1}, and Figure \ref{fig3}). This seems to be in agreement with previous studies showing that the 2008-2010 minimum exhibited peculiar properties such as cooler temperatures \cite{Issautier2008} that may explain the depleted ion abundances \cite{Lepri2013}. We therefore decide to add an additional separation in solar maximum and solar minimum periods to test for changes of the compound cross section results.

The detailed results per solar wind type and solar period for solar wind He$^{++}$ density and velocity, elemental and charge-state abundances, are given in Tables \ref{Tab1} and \ref{Tab2}. The uncertainties of the measures are determined as the dispersion (width) at half maximum of the histogram distributions used to calculate the mode values per solar wind type.

The elemental abundances (Table \ref{Tab1}) calculated for the full ACE database span for streamer and CH populations are consistent with the most recent analysis of the Ulysses data \cite<Table 1 in>[]{VonSteiger2015}, based on the same $\frac{O^{7+}}{O^{6+}} * \frac{C^{6+}}{C^{5+}}$ threshold. A detailed comparison with the analyses of \citeA{Zhao2017a} and \citeA{Zhao2022} did not seem relevant, since the streamer/CH separation is not based on the same criteria. The elemental abundances are systematically depleted for CH wind, except in the case of carbon, which is more abundant in this type of wind. For most quantities, the solar maximum values are enhanced compared to solar minimum. The change is particularly sharp in the He$^{++}$ properties (density and ratio to oxygen), as well as the $\frac{O^{7+}}{O^{6+}}$ values. Notable exceptions are the C/O and Ne/O abundances that show an increasing trend between solar maximum to solar minimum. The maximum to minimum trends we find agree with the analysis of \citeA{Lepri2013} of the ACE 1.1 database for the same periods.

Changes from maximum to minimum are also noticeable in some charge-state abundance ratios (Table \ref{Tab2}), with the highest charge states showing a decreasing trend, while the lower charge states, such as Ne$^{8+}$, and the heavier metals (Mg$^{7+,6+}$, Si$^{8+,7+}$, Fe$^{9+,8+,7+}$) seem more abundant during solar minimum, particularly for CH solar wind type. However, these trends should be considered with caution, due to the lower statistics of the ACE data during solar minimum. 

\begin{figure}
\noindent\includegraphics[width=\textwidth]{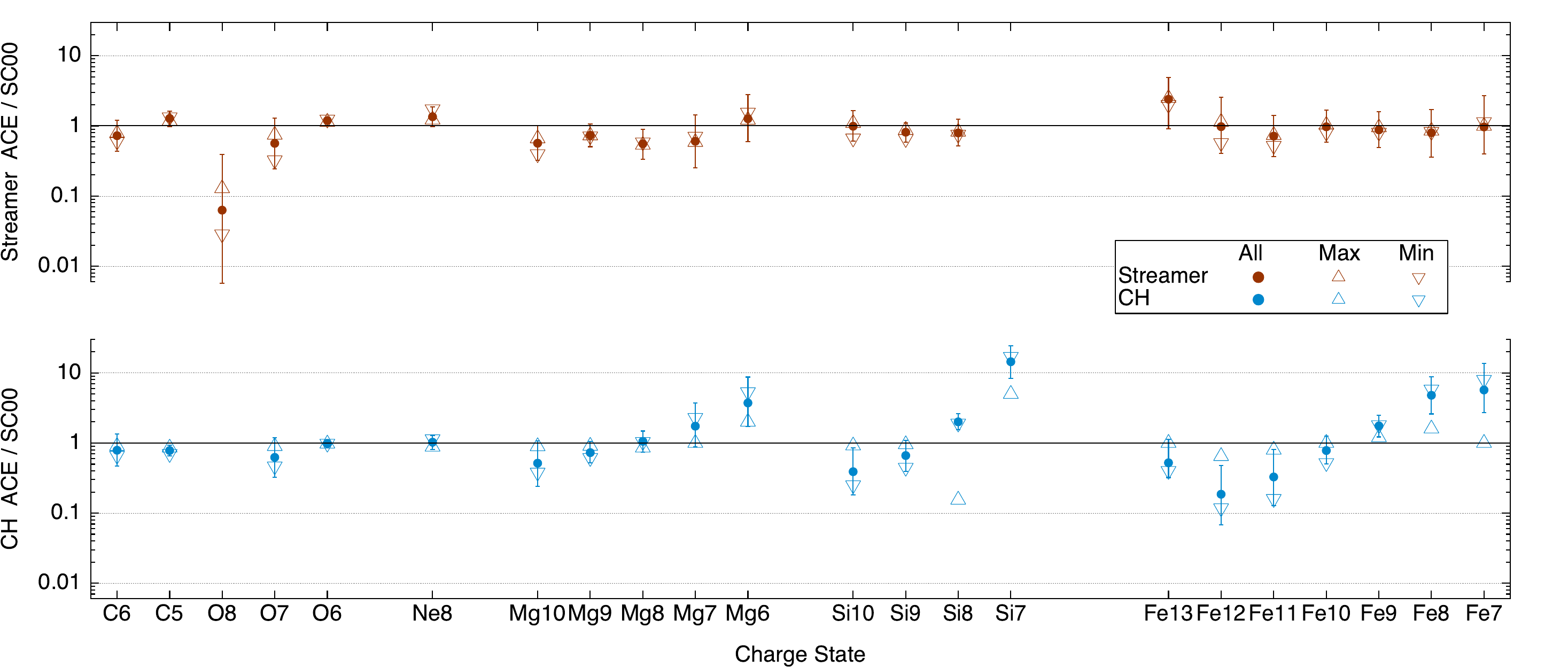}
\caption{Ratio of the ACE 1.1 / SC00 charge-state abundances for the complete 1998-2011 period (full circles), the solar maximum period (upward triangles) and solar minimum period (downward triangles), for the streamer (top panel - dark red) and CH (bottom panel - blue) solar wind types. Error-bars are provided for the complete database values only so as to not overcrowd the plot.} 
\label{figACEvsSC}
\end{figure}

In Figure \ref{figACEvsSC} we plot the ACE 1.1 to the SC00 charge state abundance ratio for the streamer (slow) and CH (fast) populations, for the complete database, as well as for the maximum and minimum periods. We also plot indicatively the error-bars for the complete database values (we omit the error-bars for the solar maximum and minimum periods so as to not overcrowd the plot). SC00 did not provide any uncertainties, thus the plotted error-bars are proportional to the uncertainties of the ACE 1.1 charge state abundances from Table \ref{Tab2}, and showcase the scatter of the measures. In general the ACE data are somewhat lower than the SC00 Ulysses values, except for the lower charge states of Mg, Si and Fe in the CH wind. The ACE data during solar maximum are a closer match to the SC00 values, especially for the same heavier metals (Mg, Si, Fe) in the CH type wind. This may be due to the fact that the Ulysses data used in SC00 spanned the period around solar maximum. It should be noted though, that the slow and fast selection in the SC00 study was based on a velocity threshold in contrast to the present analysis were we use the $\frac{O^{7+}}{O^{6+}} * \frac{C^{6+}}{C^{5+}}$ product. It would have been interesting to compare the 2008-2010 charge state abundances from Ulysses with the corresponding ACE values, but unfortunately the detailed charge-state measurements in the Ulysses final archive are not available.

\begin{sidewaystable}
 \caption{Alpha particle parameters and element abundance ratios from ACE 1.1 database for streamer (Slow), CH (fast) and ICME solar wind for different solar activity periods. The uncertainties represent the dispersion of the measures at half maximum of the distribution.}
 \centering
 \begin{tabular}{l c c c c c c c c c}
 \hline
  &\multicolumn{3}{c}{Full period (1998-2011)}&\multicolumn{3}{c}{Max (2000-2002)}&\multicolumn{3}{c}{Min (2008-2010)}\\
  & Str.   &CH   &ICME    & Str.   &CH   &ICME    & Str.   &CH   &ICME  \\
 \hline
n$_{He}$ (cm$^{-3}$)& 0.148  $^{+0.226}_{-0.091}$ &0.122$^{+0.113}_{-0.058}$  &0.166  $^{+0.328}_{-0.112}$& 0.189$^{+  0.203}_{-  0.099}$ & 0.157$^{+  0.127}_{-  0.068}$ & 0.179$^{+  0.315}_{-  0.112}$
&0.059$^{+  0.128}_{-  0.041}$ & 0.081$^{+  0.097}_{-  0.044}$ &0.057$^{+  0.099}_{-  0.036}$\\

V$_{He}$ (km~s$^{-1}$)& 384.7$^{+114.0}_{-88}$ & 610.5$^{+  107.6}_{-  91.5}$ &422.3  $^{+118.1}_{-92.3}$ &396.9 $^{+104.0}_{-82.4}$ &634. $^{+  136.0}_{-112.0}$ & 429.9$^{+125.5}_{-97.1}$ &340.7$^{+70.6}_{-58.5}$ &551.3$^{+127.7}_{-103.7}$ &345.1$^{+54.4}_{-45.0}$\\
 \hline
He/O    & 87.4 $^{+34.0}_{-24.1}$  &80.0$^{+16.6}_{-13.7}$   &86.3$^{+51.3}_{-32.2}$ & 99.4$^{+28.1}_{-21.9}$ & 81.0$^{+12.0}_{-10.5}$ & 96.0$^{+45.4}_{-30.8}$ &49.9$^{+37.7}_{-21.5}$& 74.8$^{+24.1}_{-18.2}$ & 52.8 $^{+24.8}_{-16.9}$\\

C/O     & 0.620$^{+0.078}_{-0.069}$ &0.651$^{+0.055}_{-0.051}$ &0.599$^{+0.144}_{-0.116}$ &0.615$^{+0.078}_{-0.069}$ &0.642$^{+0.040}_{-0.038}$ &0.592$^{+0.140}_{-0.113}$ &0.626$^{+0.092}_{-0.081}$ &0.635$^{+0.069}_{-0.063}$ &0.608 $^{+0.110}_{-0.093}$\\

Ne/O    & 0.111$^{+0.041}_{-0.030}$  &0.103$^{+0.028}_{-0.022}$   &0.127$^{+0.054}_{-0.038}$ 
&0.104$^{+0.034}_{-0.026}$ &0.090$^{+0.022}_{-0.017}$ &0.122$^{+0.053}_{-0.037}$
&0.142$^{+0.052}_{-0.038}$ &0.117$^{+0.024}_{-0.020}$ &0.142$^{+0.039}_{-0.031}$\\

Mg/O    & 0.134$^{+0.052}_{-0.037}$  &0.099$^{+0.023}_{-0.019}$   &0.159$^{+0.078}_{-0.052}$ 
&0.147$^{+0.059}_{-0.042}$ &0.104$^{+0.023}_{-0.019}$ &0.170$^{+0.075}_{-0.052}$
&0.125$^{+0.048}_{-0.035}$ &0.099$^{+0.027}_{-0.021}$ &0.128$^{+0.049}_{-0.035}$\\

Si/O    & 0.151$^{+0.052}_{-0.039}$  &0.123$^{+0.023}_{-0.020}$   &0.169$^{+0.066}_{-0.047}$ 
&0.172$^{+0.060}_{-0.045}$ &0.123$^{+0.032}_{-0.026}$ &0.185$^{+0.067}_{-0.049}$
&0.133$^{+0.044}_{-0.033}$ &0.119$^{+0.024}_{-0.020}$ &0.132$^{+0.052}_{-0.037}$\\

Fe/O    & 0.121$^{+0.071}_{-0.045}$  &0.092$^{+0.024}_{-0.019}$   &0.127$^{+0.096}_{-0.055}$ 
&0.136$^{+0.077}_{-0.049}$ &0.096$^{+0.035}_{-0.026}$ &0.145$^{+0.107}_{-0.062}$
&0.108$^{+0.067}_{-0.041}$ &0.092 $^{+0.025}_{-0.019}$ &0.109$^{+0.066}_{-0.041}$\\
    \hline
O$^{7+}$/O$^{6+}$ &0.131$^{+0.210}_{-0.080}$ &0.020$^{+0.017}_{-0.009}$ &0.317$^{+0.585}_{-0.204}$
&0.180$^{+0.291}_{-0.110}$ &0.029$^{+0.010}_{-0.007}$ &0.353 $^{+0.548}_{-0.218}$
&0.073$^{+0.069}_{-0.036}$ &0.015$^{+0.016}_{-0.008}$ &0.103$^{+0.068}_{-0.038}$\\

C$^{6+}$/C$^{5+}$ &0.859$^{+0.876}_{-0.420}$ &0.187$^{+0.124}_{-0.076}$ &1.291$^{+1.560}_{-0.718}$
&0.990$^{+0.956}_{-0.498}$ &0.196$^{+0.071}_{-0.051}$ &1.354$^{+1.497}_{-0.711}$
&0.654$^{+  0.443}_{-0.263}$ &0.176$^{+0.160}_{-0.081}$ &0.797$^{+0.433}_{-0.286}$\\
    \hline
 \end{tabular}
\label{Tab1} 
\end{sidewaystable}

\begin{landscape}
\begin{longtable}[c]{l c c c c c c c c c}
\caption{Charge state abundance ratios relative to oxygen $\frac{X^{q+}}{O}$ from ACE 1.1 database for streamer (Slow), CH (fast) and ICME solar wind for different solar activity periods.The uncertainties represent the dispersion of the measures at half maximum of the distribution.}\\  \hline

  &\multicolumn{3}{c}{Full period (1998-2011)}&\multicolumn{3}{c}{Max (2000-2002)}& \multicolumn{3}{c}{Min (2008-2010)}\\
Ion &Str. &CH &ICME &Str. &CH &ICME &Str. &CH &ICME \\
 \hline
C$^{6+}$ &0.231$^{+0.151}_{-0.092}$ &0.067$^{+0.047}_{-0.027}$ &0.291  $^{+0.232}_{-0.130}$  
&0.250$^{+0.1604}_{-0.095}$ &0.077 $^{+0.026}_{-0.021}$ &0.296$^{+0.221}_{-0.126}$
&0.191$^{+0.120}_{-0.068}$ &0.056$^{+0.051}_{-0.027}$ &0.215$^{+0.126}_{-0.074}$\\
C$^{5+}$ &0.269$^{+0.076}_{-0.064}$ &0.344$^{+0.066}_{-0.054}$ &0.221$^{+0.100}_{-0.071}$
&0.248$^{+0.085}_{-0.064}$ &0.377$^{+0.048}_{-0.043}$ &0.207$^{+0.094}_{-0.062}$ 
&0.280$^{+0.053}_{-0.044}$ &0.308$^{+0.049}_{-0.047}$ &0.270$^{+0.052}_{-0.042}$\\

N$^{7+}$$^{(a)}$&0.006  &0.000  \\
N$^{6+}$$^{(a)}$&0.058  &0.011  \\
N$^{5+}$$^{(a)}$&0.065  &0.127  \\

O$^{8+}$ &0.004$^{+0.023}_{-0.004}$ &$<$0.001 &0.028$^{+0.191}_{-0.025}$ & 0.009$^{+0.051}_{-0.007}$& $<$0.001& 0.037$^{+0.178}_{-0.031}$& 0.002$^{+0.005}_{-0.001}$& $<$0.001 &0.002$^{+0.007}_{-0.002}$\\
O$^{7+}$ &0.113$^{+0.145}_{-0.064}$ &0.019$^{+0.017}_{-0.009}$ &0.237$^{+0.174}_{-0.102}$ & 0.149$^{+0.162}_{-0.080}$& 0.027$^{+0.010}_{-0.008}$& 0.245$^{+0.170}_{-0.098}$& 0.065$^{+0.058}_{-0.030}$& 0.014$^{+0.015}_{-0.007}$ &0.088$^{+0.050}_{-0.033}$\\
O$^{6+}$ &0.874$^{+0.070}_{-0.071}$ &0.953$^{+0.013}_{-0.009}$ &0.764$^{+0.180}_{-0.142}$ & 0.832 $^{+0.111}_{-0.101}$& 0.946$^{+0.020}_{-0.002}$& 0.727$^{+0.175}_{-0.146}$& 0.904$^{+0.040}_{-0.044}$& 0.952$^{+0.014}_{-0.009}$ &0.887$^{+0.056}_{-0.047}$\\
O$^{5+}$ &0.019$^{+  0.009}_{-  0.006}$	&0.025$^{+  0.009}_{-  0.007}$ &0.016$^{+  0.010}_{-  0.006}$ & 0.017$^{+  0.008}_{-  0.005}$& 0.020$^{+  0.006}_{-  0.005}$& 0.015$^{+  0.008}_{-  0.005}$& 0.024$^{+  0.012}_{-  0.008}$& 0.027 $^{+  0.010}_{-  0.007}$ &0.024$^{+  0.012}_{-  0.008}$\\

Ne$^{9+}$ &0.003$^{+  0.007}_{-  0.002}$ &0.001$^{+  0.002}_{-  0.001}$ &0.008$^{+  0.033}_{-  0.007}$ & 0.003$^{+  0.008}_{-  0.002}$& 0.001 $^{+  0.001}_{-  0.001}$& 0.008$^{+  0.033}_{-  0.007}$& 0.003$^{+  0.006}_{-  0.002}$& 0.001$^{+  0.002}_{-  0.001}$&0.003$^{+  0.006}_{-  0.002}$\\
Ne$^{8+}$ &0.114$^{+  0.043}_{-  0.032}$ &0.105$^{+  0.026}_{-  0.023}$ &0.128$^{+  0.044}_{-  0.034}$ & 0.103 $^{+  0.032}_{-  0.026}$& 0.09$^{+  0.022}_{-  0.016}$& 0.122$^{+  0.048}_{-  0.033}$& 0.146$^{+  0.059}_{-  0.039}$& 0.116$^{+  0.025}_{-  0.019}$&0.142$^{+  0.044}_{-  0.030}$\\

Mg$^{12+}$ &$<$0.001 &0.000 &0.001$^{+  0.002}_{-  0.001}$ & $<$0.001& 0.000& 0.001$^{+  0.003}_{-  0.001}$& 0.000& 0.000&0.001$^{+  0.001}_{-  0.001}$\\
Mg$^{11+}$ &0.001$^{+  0.001}_{-  0.001}$ &$<$0.001 &0.001$^{+  0.008}_{-  0.001}$ & 0.001$^{+  0.001}_{-  0.001}$& 0.000& 0.001$^{+  0.009}_{-  0.001}$& 0.001$^{+  0.001}_{-  0.001}$& $<$0.001&0.001$^{+  0.001}_{-  0.001}$\\
Mg$^{10+}$ &0.055$^{+  0.043}_{-  0.024}$ &0.015$^{+  0.016}_{-  0.008}$ &0.088$^{+  0.078}_{-  0.042}$ & 0.065$^{+  0.043}_{-  0.026}$& 0.026$^{+  0.011}_{-  0.008}$& 0.095$^{+  0.076}_{-  0.043}$& 0.039$^{+  0.035}_{-  0.018}$& 0.011$^{+  0.011}_{-  0.005}$&0.046$^{+  0.035}_{-  0.020}$\\
Mg$^{9+}$ &0.038$^{+  0.017}_{-  0.012}$ &0.032$^{+  0.014}_{-  0.009}$ &0.040$^{+  0.018}_{-  0.013}$ & 0.038$^{+  0.017}_{-  0.012}$& 0.040$^{+  0.012}_{-  0.011}$& 0.039$^{+  0.019}_{-  0.012}$& 0.037 $^{+  0.021}_{-  0.014}$& 0.027$^{+  0.014}_{-  0.009}$&0.038$^{+  0.024}_{-  0.015}$\\
Mg$^{8+}$ &0.023$^{+  0.014}_{-  0.009}$ &0.030$^{+  0.012}_{-  0.009}$ &0.020$^{+  0.014}_{-  0.008}$ & 0.022$^{+  0.014}_{-  0.009}$& 0.023$^{+  0.011}_{-  0.008}$& 0.020$^{+  0.014}_{-  0.008}$& 0.024$^{+  0.017}_{-  0.009}$& 0.029$^{+  0.012}_{-  0.008}$&0.023$^{+  0.019}_{-  0.010}$\\
Mg$^{7+}$ &0.010$^{+  0.014}_{-  0.006}$ &0.012$^{+  0.014}_{-  0.006}$ &0.009$^{+  0.012}_{-  0.005}$ & 0.010$^{+  0.013}_{-  0.006}$& 0.007$^{+  0.009}_{-  0.004}$& 0.008$^{+  0.010}_{-  0.005}$& 0.012$^{+  0.017}_{-  0.007}$& 0.016$^{+  0.014}_{-  0.007}$&0.013$^{+  0.016}_{-  0.007}$\\
Mg$^{6+}$ &0.011$^{+  0.014}_{-  0.006}$ &0.011$^{+  0.015}_{-  0.006}$ &0.009$^{+  0.012}_{-  0.005}$   & 0.011$^{+  0.012}_{-  0.006}$& 0.006$^{+  0.009}_{-  0.004}$& 0.008$^{+  0.010}_{-  0.005}$& 0.014$^{+  0.018}_{-  0.008}$& 0.016$^{+  0.015}_{-  0.008}$&0.015$^{+  0.019}_{-  0.009}$\\

Si$^{12+}$ &0.006$^{+  0.015}_{-  0.004}$  &0.001$^{+  0.002}_{-  0.001}$ &0.021$^{+  0.065}_{-  0.016}$ & 0.008$^{+  0.021}_{-  0.006}$& 0.001$^{+  0.002}_{-  0.001}$& 0.025$^{+  0.074}_{-  0.018}$& 0.003$^{+  0.005}_{-  0.002}$& 0.001$^{+  0.001}_{-  0.001}$&0.003$^{+  0.008}_{-  0.002}$\\
Si$^{11+}$ &0.016$^{+  0.016}_{-  0.008}$  &0.005$^{+  0.005}_{-  0.002}$ &0.029$^{+  0.035}_{-  0.016}$ & 0.019$^{+  0.017}_{-  0.009}$& 0.008$^{+  0.006}_{-  0.003}$& 0.031$^{+  0.034}_{-  0.016}$& 0.010$^{+  0.011}_{-  0.005}$& 0.003$^{+  0.004}_{-  0.002}$&0.011$^{+  0.012}_{-  0.006}$\\
Si$^{10+}$ &0.021$^{+  0.014}_{-  0.008}$  &0.010$^{+  0.011}_{-  0.005}$ &0.025$^{+  0.014}_{-  0.009}$ & 0.023$^{+  0.011}_{-  0.008}$& 0.022$^{+  0.009}_{-  0.006}$& 0.025$^{+  0.012}_{-  0.009}$& 0.014$^{+  0.013}_{-  0.007}$& 0.006$^{+  0.005}_{-  0.003}$&0.016$^{+  0.012}_{-  0.007}$\\
Si$^{9+}$ &0.040$^{+  0.014}_{-  0.011}$  &0.030$^{+  0.019}_{-  0.012}$ &0.039$^{+  0.019}_{-  0.013}$ & 0.043$^{+  0.013}_{-  0.010}$& 0.043$^{+  0.013}_{-  0.010}$& 0.040$^{+  0.018}_{-  0.012}$& 0.032$^{+  0.017}_{-  0.011}$& 0.020$^{+  0.010}_{-  0.006}$&0.031$^{+  0.016}_{-  0.010}$\\
Si$^{8+}$ &0.045$^{+  0.026}_{-  0.016}$  &0.044$^{+  0.014}_{-  0.010}$ &0.038$^{+  0.031}_{-  0.017}$ & 0.047$^{+  0.030}_{-  0.018}$& 0.034$^{+  0.017}_{-  0.010}$& 0.038$^{+  0.034}_{-  0.018}$& 0.043$^{+  0.021}_{-  0.014}$& 0.042$^{+  0.012}_{-  0.010}$&0.043$^{+  0.022}_{-  0.014}$\\
Si$^{7+}$ &0.022$^{+  0.023}_{-  0.012}$  &0.029$^{+  0.020}_{-  0.012}$ &0.015$^{+  0.024}_{-  0.009}$ & 0.021$^{+  0.025}_{-  0.011}$& 0.010$^{+  0.009}_{-  0.005}$& 0.016$^{+  0.023}_{-  0.009}$& 0.027$^{+  0.024}_{-  0.012}$& 0.034$^{+  0.016}_{-  0.011}$&0.026$^{+  0.022}_{-  0.012}$\\
Si$^{6+}$ &0.004$^{+  0.008}_{-  0.003}$  &0.006$^{+  0.011}_{-  0.004}$ &0.003$^{+  0.006}_{-  0.002}$ & 0.004$^{+  0.007}_{-  0.002}$& 0.002$^{+  0.003}_{-  0.001}$& 0.003$^{+  0.005}_{-  0.002}$& 0.006$^{+  0.008}_{-  0.003}$& 0.010$^{+  0.009}_{-  0.004}$&0.006$^{+  0.009}_{-  0.004}$\\

S$^{11+}$$^{(a)}$&0.000 &0.001  & & & & & & & \\
S$^{10+}$$^{(a)}$&0.005 &0.008  & & & & & & & \\
S$^{9+}$$^{(a)}$&0.016  &0.027  & & & & & & & \\
S$^{8+}$$^{(a)}$&0.019  &0.023  & & & & & & & \\
S$^{7+}$$^{(a)}$&0.006  &0.005  & & & & & & & \\

Fe$^{20+}$ &$<$0.001  &0.000	&$<$0.001  & $<$0.001& 0.000& $<$0.001& $<$0.001& 0.000&$<$0.001\\
Fe$^{19+}$ &$<$0.001  &$<$0.001	&0.001$^{+  0.002}_{-  0.001}$  & $<$0.001& $<$0.001& 0.001$^{+  0.002}_{-  0.001}$& $<$0.001& $<$0.001&$<$0.001\\
Fe$^{18+}$ &$<$0.001  &$<$0.001	&0.001$^{+  0.002}_{-  0.001}$  & $<$0.001& $<$0.001& 0.001$^{+  0.003}_{-  0.001}$& 0.001$^{+  0.001}_{-  0.001}$& $<$0.001&0.001$^{+  0.001}_{-  0.001}$\\
Fe$^{17+}$ &0.001$^{+  0.001}_{-  0.001}$  &$<$0.001	&0.002$^{+  0.006}_{-  0.001}$  & 0.001$^{+  0.001}_{-  0.001}$& $<$0.001& 0.002$^{+  0.007}_{-  0.001}$& 0.001$^{+  0.001}_{-  0.001}$& $<$0.001&0.001$^{+  0.001}_{-  0.001}$\\
Fe$^{16+}$ &0.001$^{+  0.003}_{-  0.001}$  &0.001$^{+  0.001}_{-  0.001}$	&0.006$^{+  0.043}_{-  0.005}$  & 0.002$^{+  0.005}_{-  0.001}$& 0.001$^{+  0.00}_{-  0.001}$& 0.008$^{+  0.058}_{-  0.007}$& 0.001$^{+  0.002}_{-  0.001}$& 0.001$^{+  0.001}_{-  0.001}$&0.001$^{+  0.002}_{-  0.001}$\\
Fe$^{15+}$ &0.002$^{+  0.003}_{-  0.001}$  &0.001$^{+  0.001}_{-  0.001}$	&0.005$^{+  0.015}_{-  0.004}$  & 0.002$^{+  0.004}_{-  0.001}$& 0.001$^{+  0.001}_{-  0.001}$& 0.005$^{+  0.018}_{-  0.004}$& 0.002$^{+  0.003}_{-  0.001}$& 0.001$^{+  0.001}_{-  0.001}$&0.002$^{+  0.003}_{-  0.001}$\\
Fe$^{14+}$ &0.004$^{+  0.004}_{-  0.002}$  &0.002$^{+  0.002}_{-  0.001}$	&0.006$^{+  0.009}_{-  0.004 }$  & 0.004$^{+  0.004}_{-  0.002}$& 0.003$^{+  0.002}_{-  0.001}$& 0.006$^{+  0.009}_{-  0.004}$& 0.003$^{+  0.004}_{-  0.002}$& 0.002$^{+  0.002}_{-  0.001}$&0.003$^{+  0.004}_{-  0.002}$\\
Fe$^{13+}$ &0.005$^{+  0.005}_{-  0.003}$  &0.003$^{+  0.003}_{-  0.001}$	&0.008 $^{+  0.011}_{-  0.005}$   & 0.005$^{+  0.006}_{-  0.003}$& 0.005$^{+  0.004}_{-  0.002}$& 0.007$^{+  0.012}_{-  0.005}$& 0.004$^{+  0.005}_{-  0.002}$& 0.002$^{+  0.002}_{-  0.001}$&0.004$^{+  0.005}_{-  0.002}$\\
Fe$^{12+}$ &0.007$^{+  0.011}_{-  0.004}$  &0.003$^{+  0.005}_{-  0.002}$	&0.010$^{+  0.015}_{-  0.006}$   & 0.008$^{+  0.011}_{-  0.005}$& 0.011$^{+  0.007}_{-  0.004}$& 0.010$^{+  0.014}_{-  0.006}$& 0.004$^{+  0.006}_{-  0.002}$& 0.002$^{+  0.002}_{-  0.001}$&0.004$^{+  0.006}_{-  0.002}$\\
Fe$^{11+}$ &0.016$^{+  0.016}_{-  0.008}$  &0.008$^{+  0.012}_{-  0.005}$	&0.018$^{+  0.016}_{-  0.009}$   & 0.017$^{+  0.014}_{-  0.008}$& 0.020$^{+  0.007 }_{-  0.006}$& 0.017$^{+  0.014}_{-  0.008}$& 0.012$^{+  0.019}_{-  0.007}$& 0.004$^{+  0.003}_{-  0.002}$&0.012$^{+  0.015}_{-  0.007}$\\
Fe$^{10+}$ &0.030$^{+  0.022}_{-  0.012}$  &0.020$^{+  0.012}_{-  0.007}$	&0.029$^{+  0.025}_{-  0.014}$   & 0.032$^{+  0.020}_{-  0.012}$& 0.025$^{+  0.010}_{-  0.007}$& 0.029$^{+  0.025}_{-  0.013}$& 0.025$^{+  0.025}_{-  0.013}$& 0.013$^{+  0.008}_{-  0.005}$&0.024$^{+  0.025}_{-  0.012}$\\
Fe$^{9+}$ &0.036$^{+  0.029}_{-  0.016}$  &0.026$^{+  0.011}_{-  0.008}$	&0.032$^{+  0.036}_{-  0.017}$   & 0.039$^{+  0.028}_{-  0.016}$& 0.018$^{+  0.015}_{-  0.008}$& 0.035$^{+  0.034}_{-  0.017}$& 0.033$^{+  0.027}_{-  0.015}$& 0.027$^{+  0.010}_{-  0.007}$&0.032$^{+  0.024}_{-  0.014}$\\
Fe$^{8+}$ &0.027$^{+  0.031}_{-  0.015}$  &0.024$^{+  0.020}_{-  0.011}$	&0.024$^{+  0.041}_{-  0.015}$   & 0.029$^{+  0.036}_{-  0.016}$& 0.008$^{+  0.015}_{-  0.005}$& 0.026$^{+  0.041}_{-  0.016}$& 0.028$^{+  0.028}_{-  0.014}$& 0.029$^{+  0.015}_{-  0.010}$&0.028$^{+  0.026}_{-  0.013}$\\
Fe$^{7+}$ &0.007$^{+  0.012}_{-  0.004}$  &0.006$^{+  0.008}_{-  0.003}$	&0.007$^{+  0.016}_{-  0.005}$   & 0.007$^{+  0.013}_{-  0.005}$& 0.001$^{+  0.003}_{-  0.001}$& 0.007$^{+  0.016}_{-  0.005}$& 0.008$^{+  0.013}_{-  0.005}$& 0.008 $^{+  0.006}_{-  0.004}$&0.009$^{+  0.012}_{-  0.005}$\\
Fe$^{6+}$ &0.002$^{+  0.005}_{-  0.002}$  &0.001$^{+  0.002}_{-  0.001}$	&0.002$^{+  0.005}_{-  0.002}$   & 0.002$^{+  0.005}_{-  0.002}$& 0.001$^{+  0.001}_{-  0.001}$& 0.002$^{+  0.005}_{-  0.002}$& 0.004$^{+  0.007}_{-  0.002}$& 0.002$^{+  0.002}_{-  0.001}$&0.003$^{+  0.004}_{-  0.002}$\\
\hline
\multicolumn{10}{l}{$^{(a)}$Unavailable in the ACE database. Numbers are from Table 1 in SC00.}\\ 
\label{Tab2}
\end{longtable}
\end{landscape}

\section{Atomic data}\label{SecAD}
\subsection{Velocity-dependent Cross-Sections}
For the CX cross-sections we are using the compilation provided by the KRONOS\footnote{https://www.physast.uga.edu/ugacxdb/} package at the University of Georgia \cite{Cumbee2021}. The KRONOS database includes comprehensive single electron cross-section values for many ion-neutral pairs, in particular H- and He-like charge states, and a wide range of collision energies. The calculations are based on the Multi-Channel Landau-Zener (MCLZ) approximation \cite{Lyons2017}. However, for several ion-neutral couples, other recommended data-sets based on more accurate methods, such as atomic-orbital close-coupling \cite<AOCC;>[]{Fritsch1991}, molecular-orbital close-coupling \cite<MOCC;>[]{Janev1993, Harel1998}, quantum-mechanical molecular-orbital close-coupling \cite<QMOCC;>[]{Nolte2012} and classical trajectory Monte Carlo \cite<CTMC;>[]{Abrines1966}, are also provided. The KRONOS database package has been applied successfully on both solar system and astrophysical CX spectra \cite{Mullen2017, Cumbee2016, Cumbee2017}, and is also employed in the AtomDB Charge eXchange v2.0 (ACX2) spectral model \cite<>[and http://www.atomdb.org/CX]{Smith2012}.   

In this study, we use the velocity-dependent cross sections according to the KRONOS preferred order QMOCC $\rightarrow$ MOCC $\rightarrow$ AOCC $\rightarrow$ CTMC $\rightarrow$ MCLZ when available. The velocity-dependent cross sections from KRONOS are interpolated to the ACE He$^{++}$ velocity time series. In short, this includes all the bare and He-like ions (C, N, O, Ne, Mg) and only a few Li-like, or lower charge states, of Ne, Mg, and S from table \ref{Tab2}. If the cross-sections of any ${X^{q+},M}$ couple are not available through KRONOS, we use the SC00 cross-section values.

\subsection{Emission Line Probabilities\label{SecEmLines}}
The emission line probabilities (or yields) are based on two types of calculations, as described in \citeA{Koutroumpa2006, Koutroumpa2009b} and references therein. 

For ions C, O, N, Ne, Mg the quantum yield cascades and emission line energies are calculated by \citeA{Kharchenko2005} for collisions with H and He respectively, and for the slow and fast SW regimes. These spectra have been successfully applied to cometary SWCX emission spectra for slow and fast SW velocities \cite<e.g.,>[]{Kharchenko2000, Kharchenko2001, Rigazio2002, Kharchenko2005}. 

For heavier ions (Fe, Si, S, Mg), the quantum yield database was updated using the hydrogenic approximation \cite<see details in >[]{Kharchenko2005,Koutroumpa2009b} without distinction between H and He targets, or slow and fast SW regimes. Within this model, the exact positions of emission lines are not accurate compared to real emission spectra, but the total energy budget of the cascades in a given energy range is correct. Since many of those ion lines are blended with each other in the lower energy range (0.1-0.3 keV), this defect is less important at low to moderate spectral resolution of current instruments. However, future missions such as the Line Emission Mapper proposed concept \cite<LEM;>{Kraft2022} and ESA's Athena mission \cite{Barret2020}, which will provide microcalorimeter-resolution data, will require a deep reevaluation of the spectral line ratios in the lower energy range. The complete spectral line database may be consulted in \citeA{Koutroumpa2007a}\footnote{https://tel.archives-ouvertes.fr/tel-00260160/document}.

\section{Compound Cross-section Results}\label{SecResults}
\begin{figure}
\noindent\includegraphics[width=\textwidth]{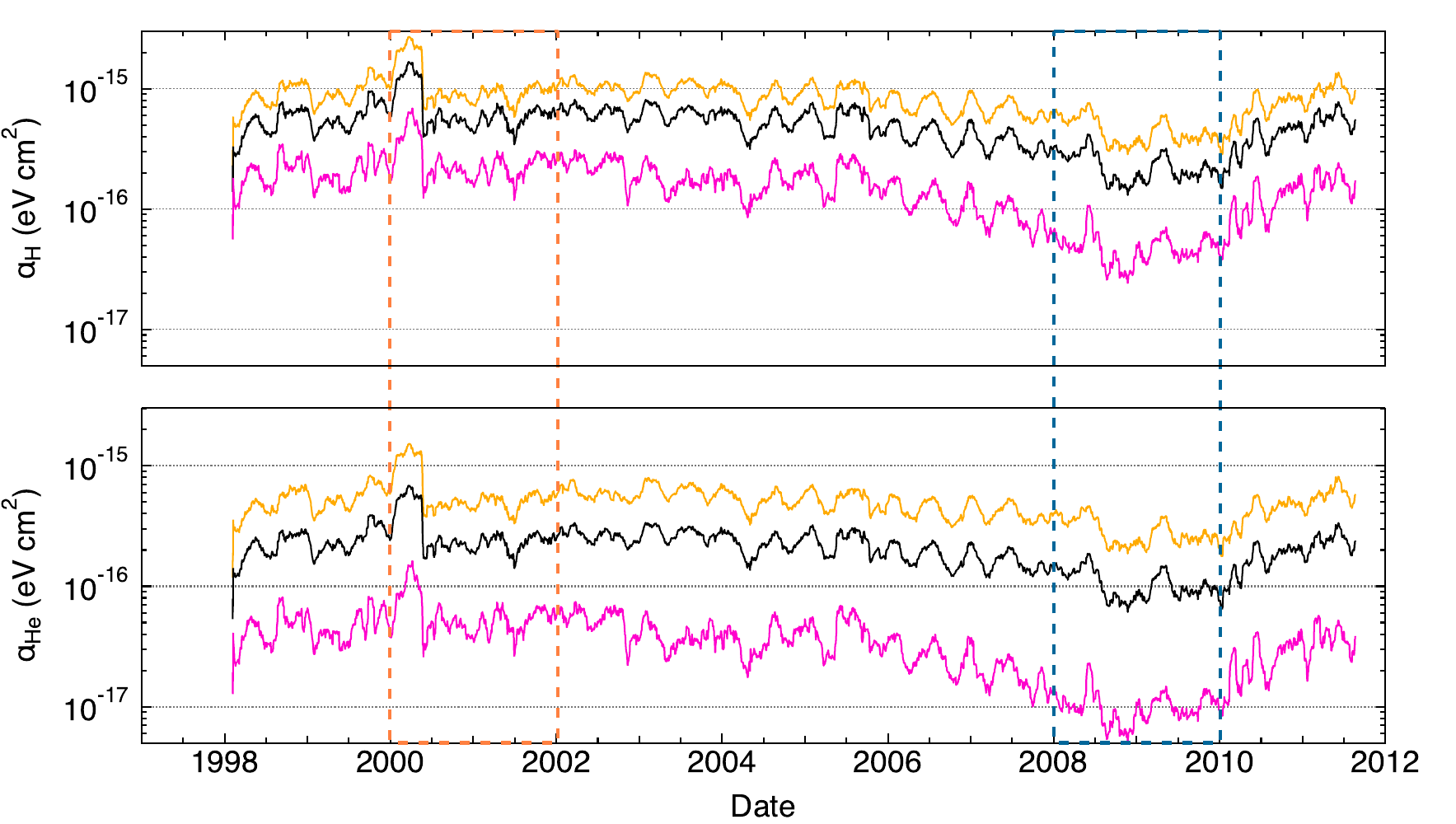}
\caption{Top- Time series of the $\alpha_H$ for the 0.1 - 2.0 keV (yellow), 0.3 - 2.0 keV (black) and 0.5 - 0.7 keV (magenta) energy ranges. Bottom- Same as the top panel except for the $\alpha_{He}$. All values are 27-day averages. The orange and blue dashed rectangles represent the solar maximum and solar minimum periods respectively. }
\label{FigAlphaTS}
\end{figure}

\begin{figure}
\noindent\includegraphics[height=3cm, width=0.24\textwidth]{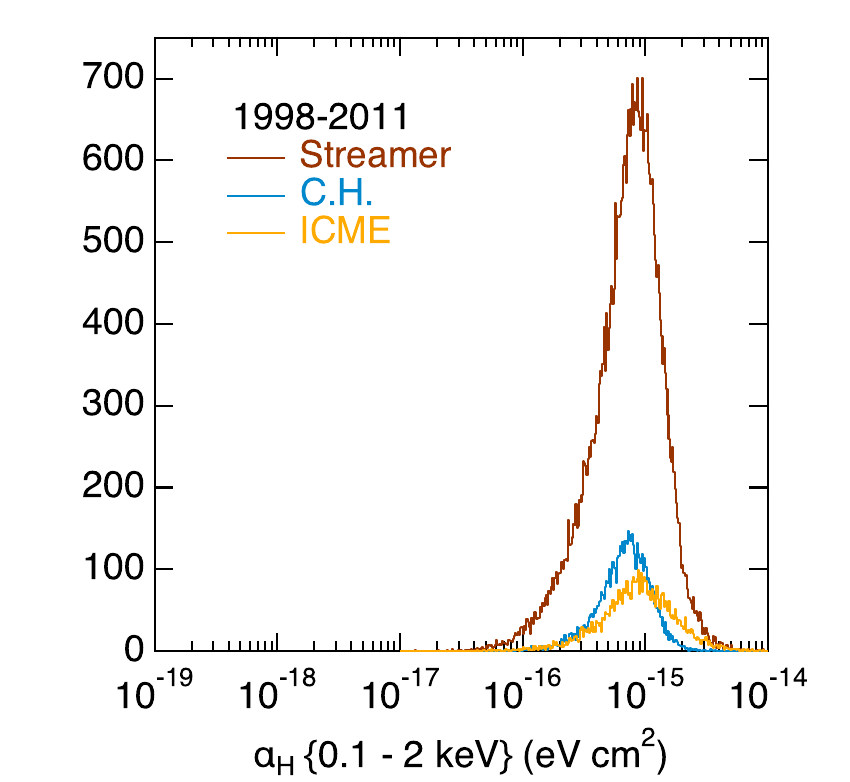}
\noindent\includegraphics[height=3cm, width=0.24\textwidth]{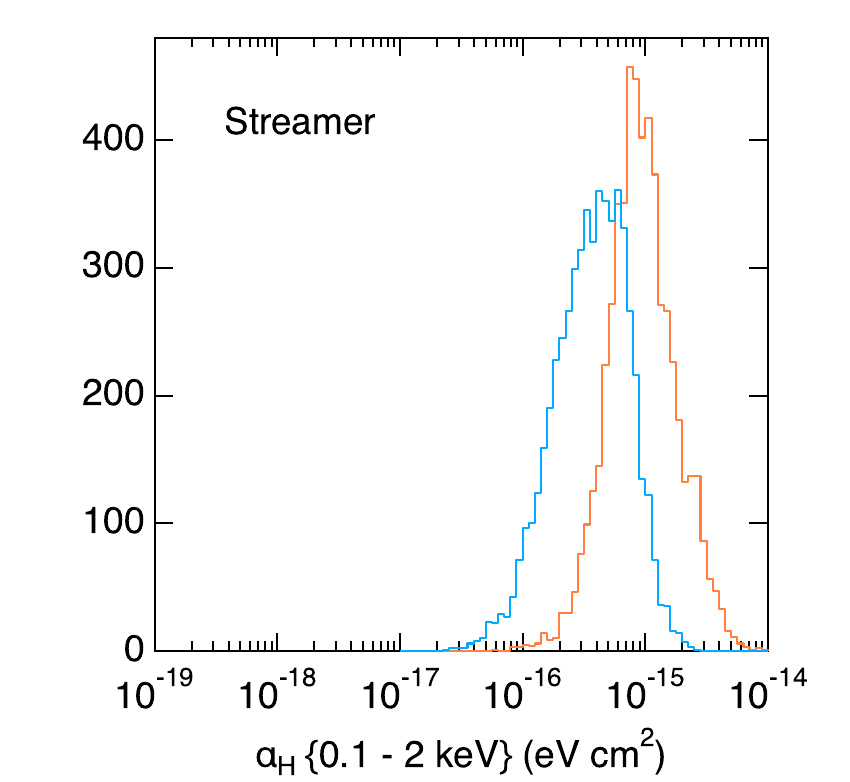}
\noindent\includegraphics[height=3cm, width=0.24\textwidth]{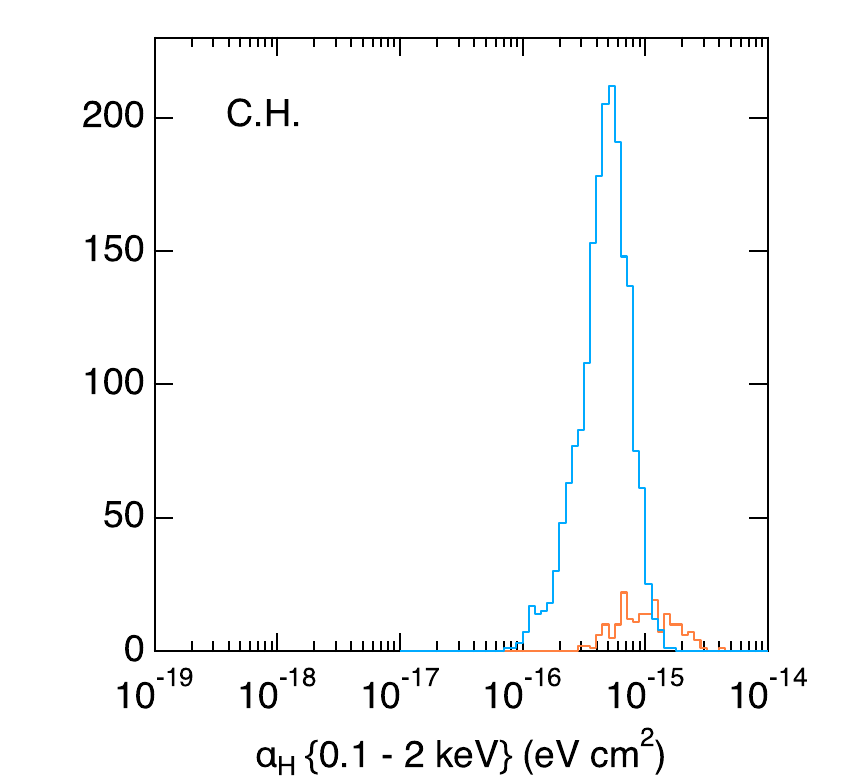}
\noindent\includegraphics[height=3cm, width=0.24\textwidth]{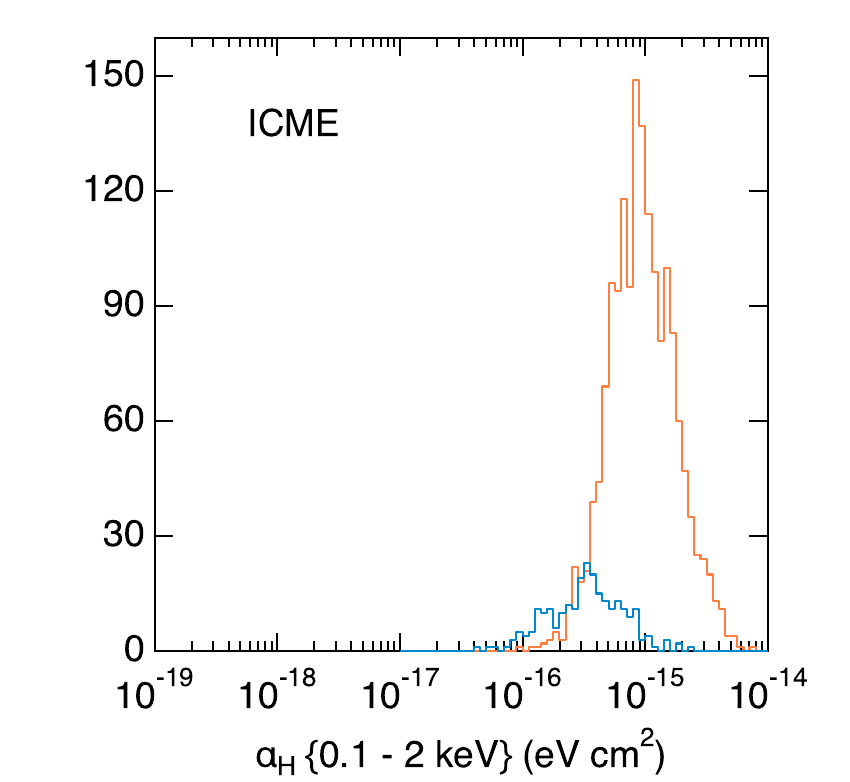}

\noindent\includegraphics[height=3cm, width=0.24\textwidth]{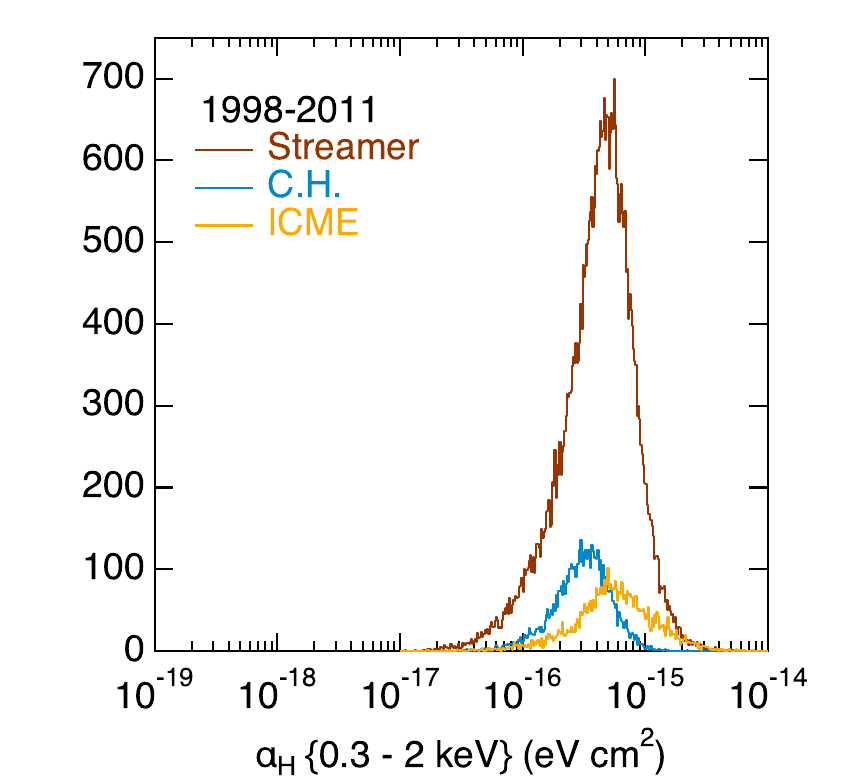}
\noindent\includegraphics[height=3cm, width=0.24\textwidth]{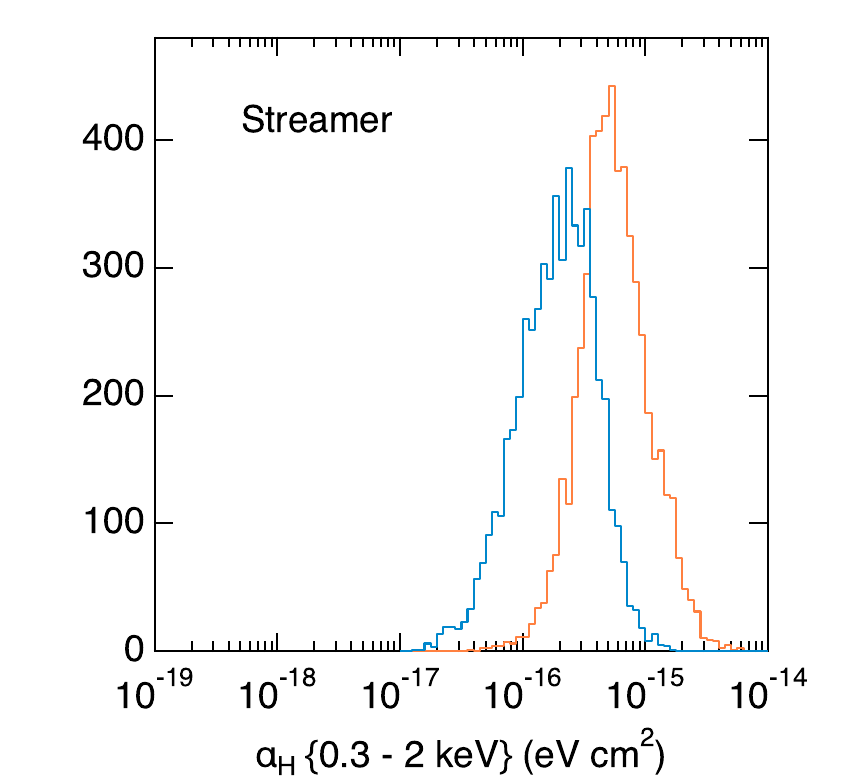}
\noindent\includegraphics[height=3cm, width=0.24\textwidth]{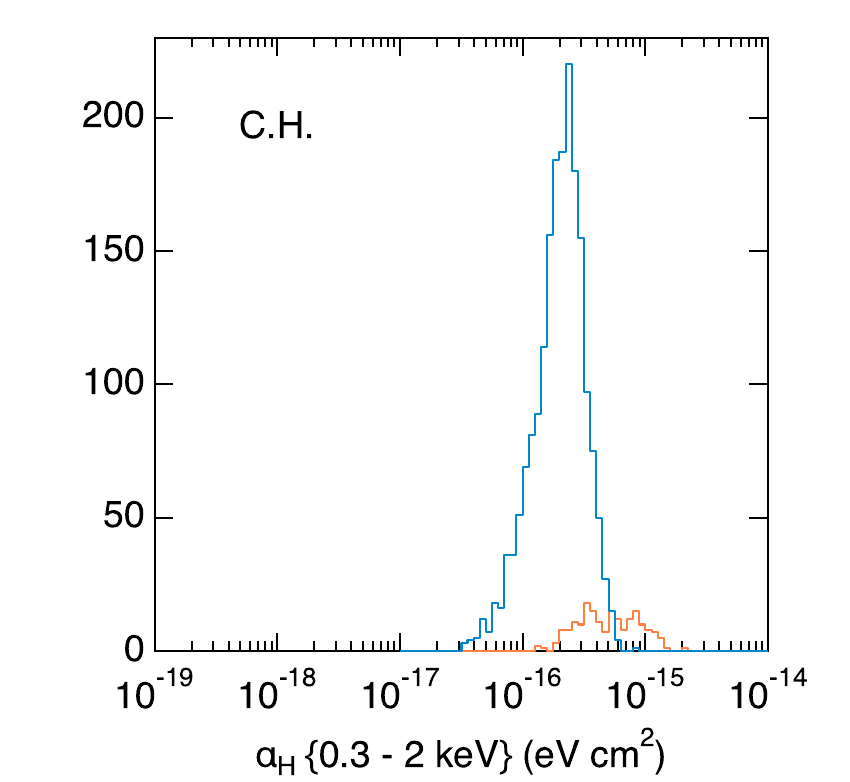}
\noindent\includegraphics[height=3cm, width=0.24\textwidth]{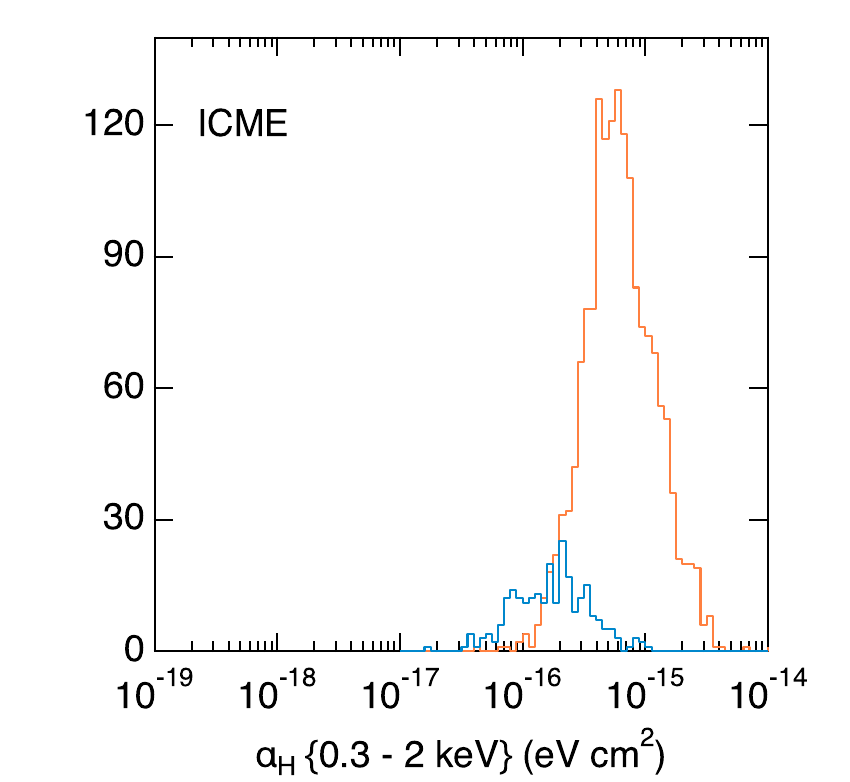}

\noindent\includegraphics[height=3cm, width=0.24\textwidth]{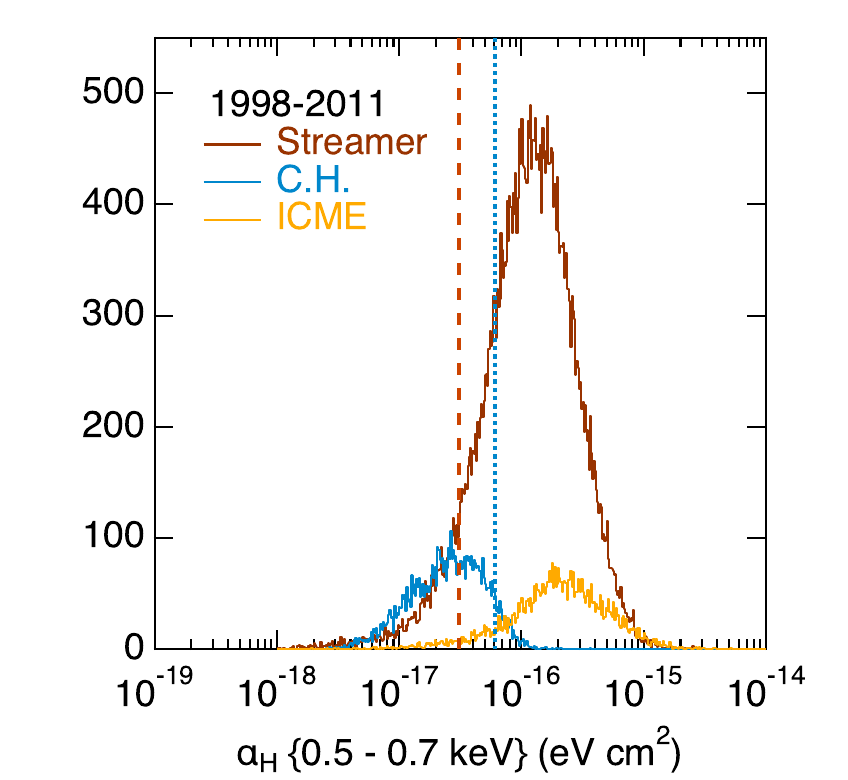}
\noindent\includegraphics[height=3cm, width=0.24\textwidth]{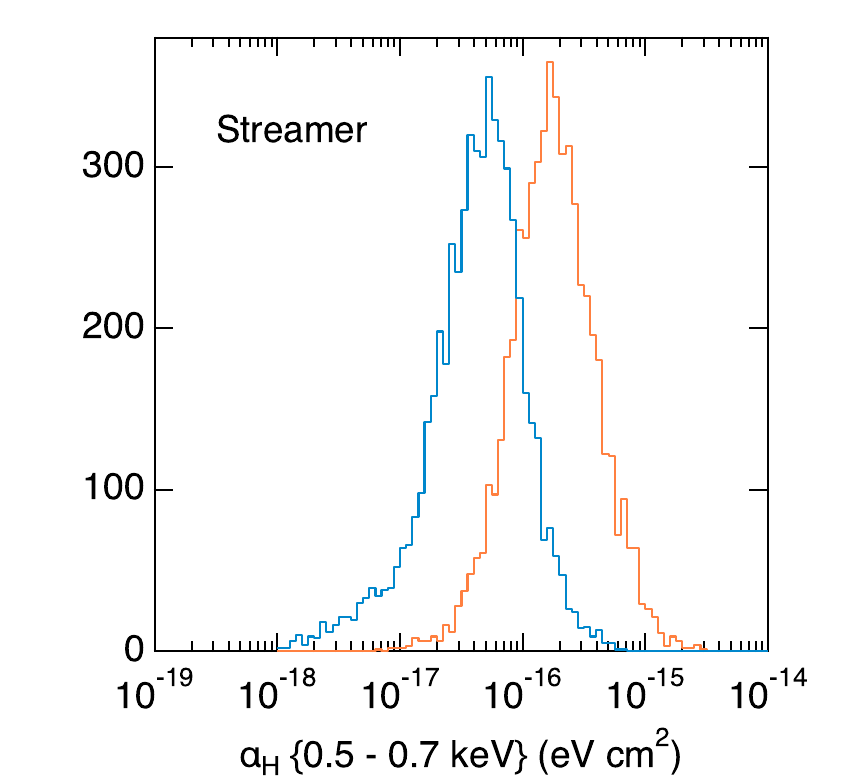}
\noindent\includegraphics[height=3cm, width=0.24\textwidth]{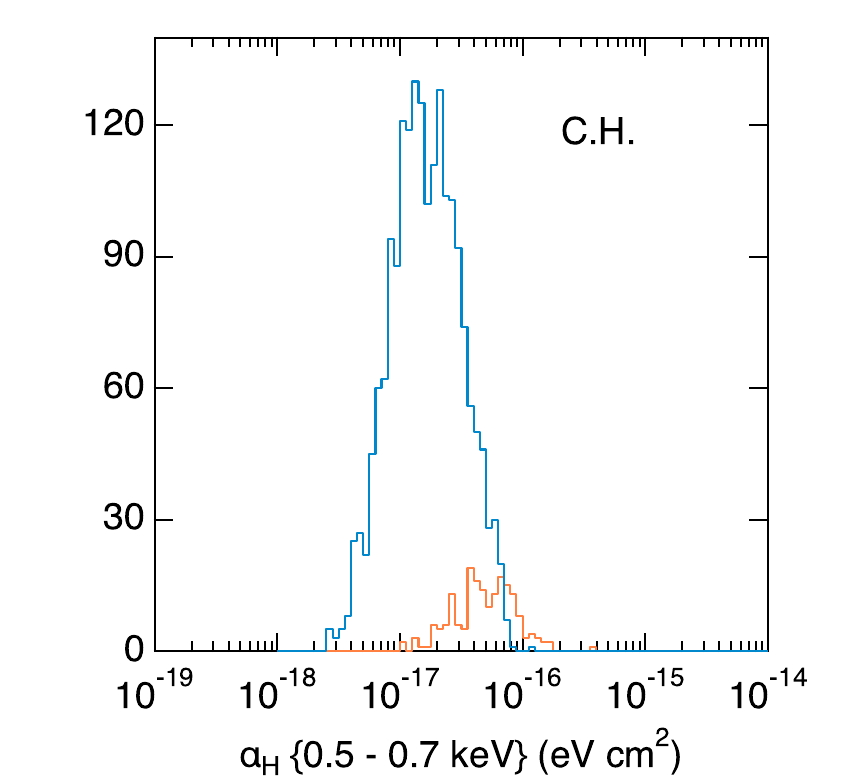}
\noindent\includegraphics[height=3cm, width=0.24\textwidth]{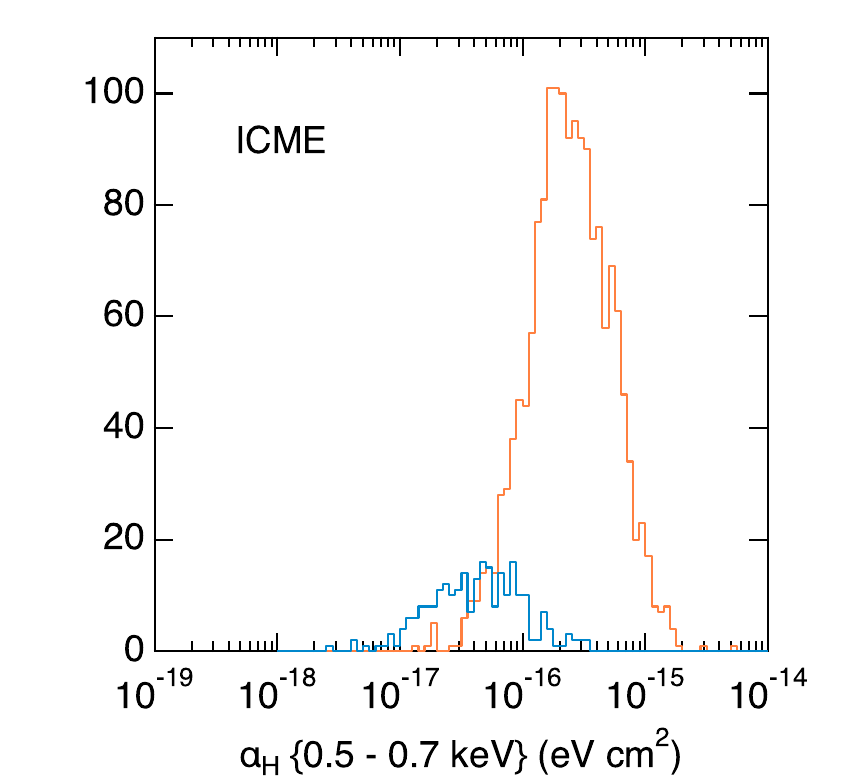}

\noindent\includegraphics[height=3cm, width=0.24\textwidth]{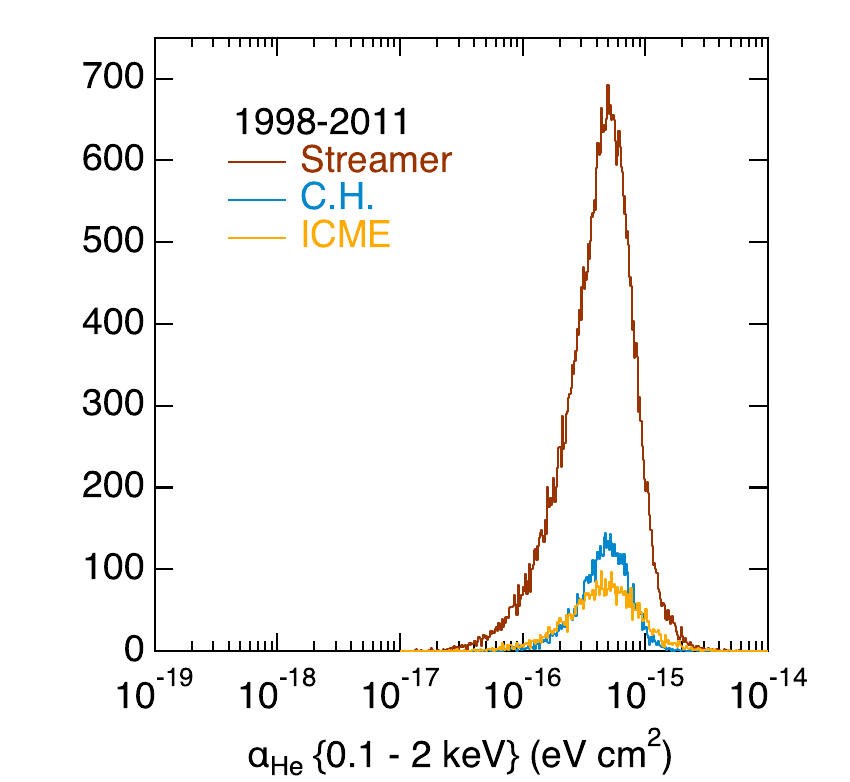}
\noindent\includegraphics[height=3cm, width=0.24\textwidth]{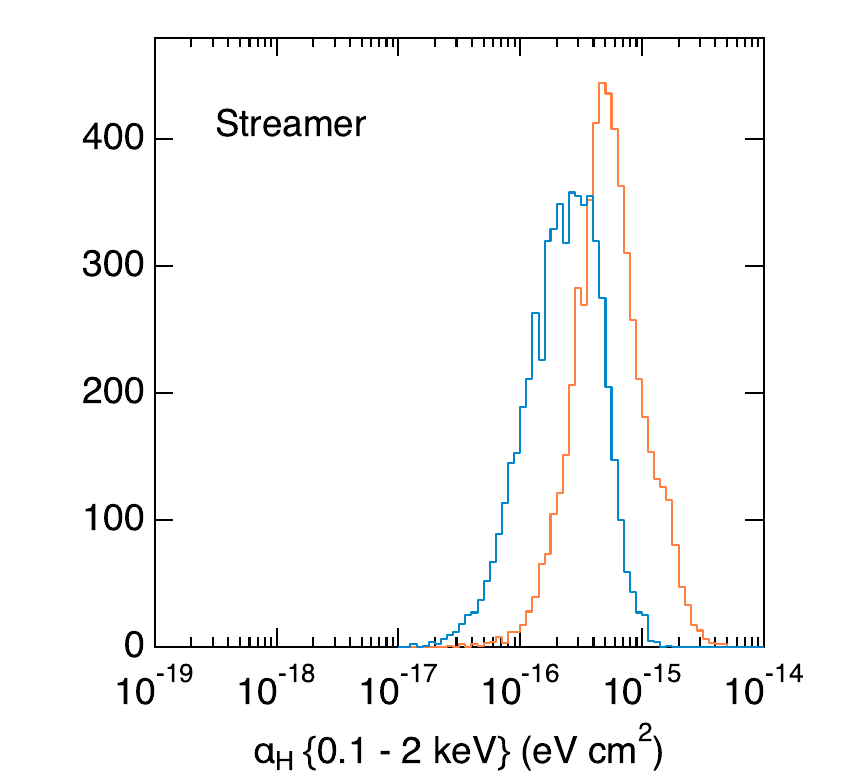}
\noindent\includegraphics[height=3cm, width=0.24\textwidth]{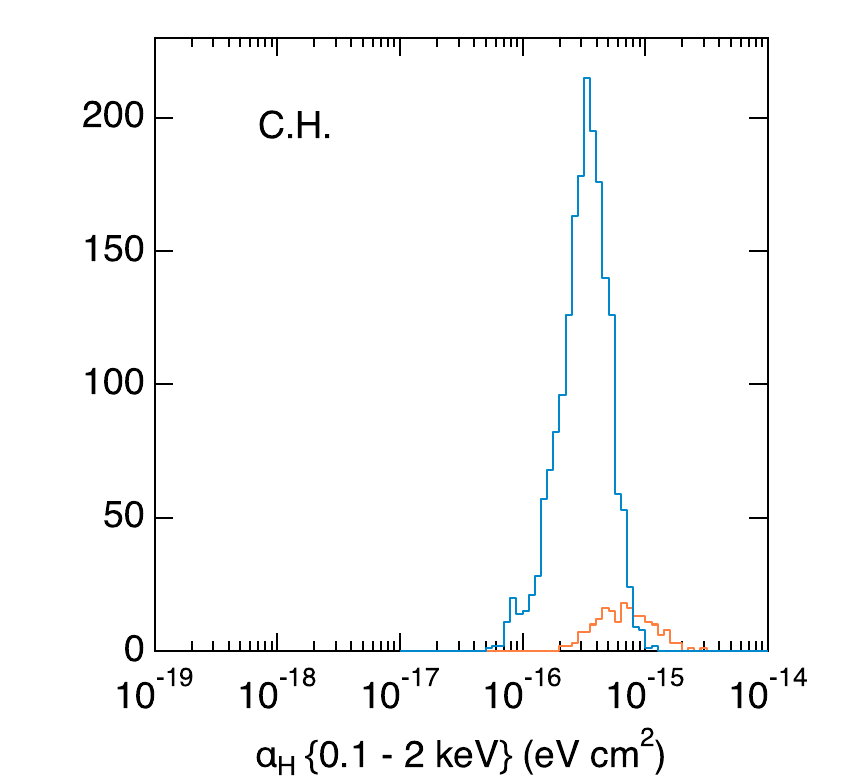}
\noindent\includegraphics[height=3cm, width=0.24\textwidth]{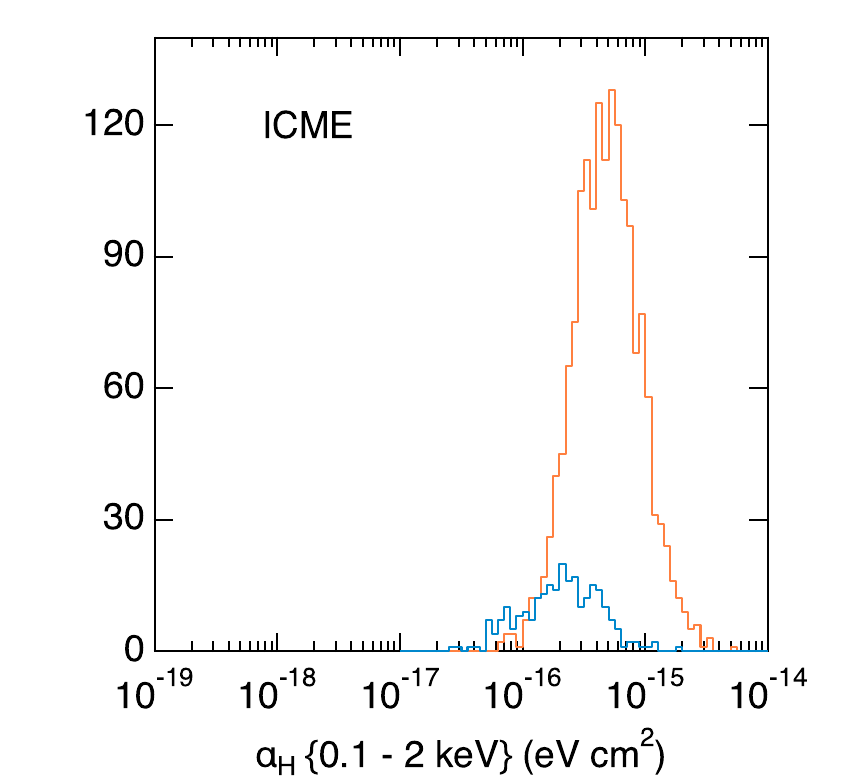}

\noindent\includegraphics[height=3cm, width=0.24\textwidth]{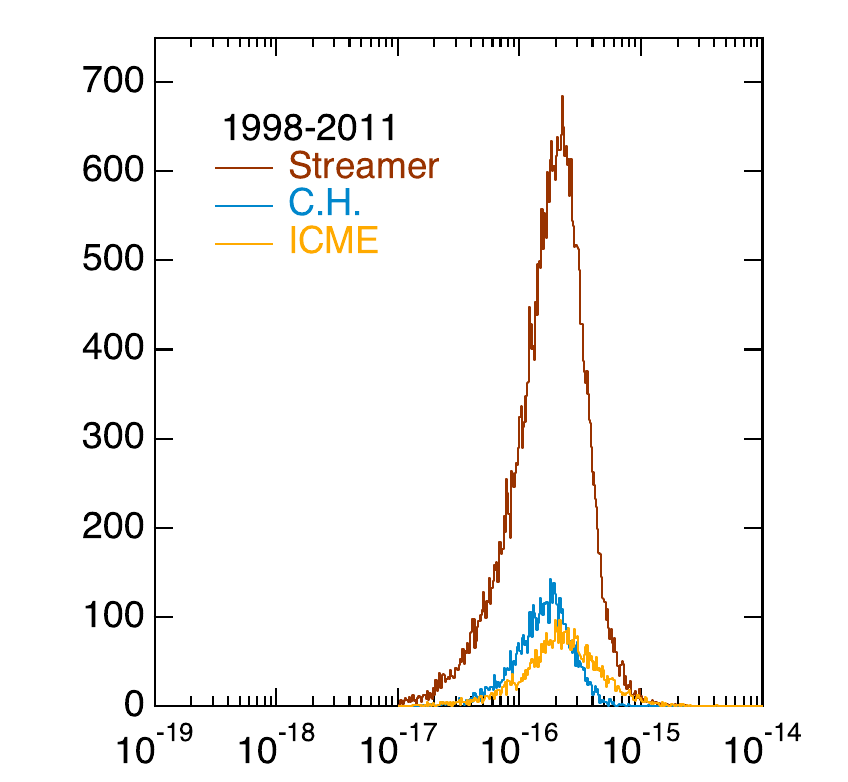}
\noindent\includegraphics[height=3cm, width=0.24\textwidth]{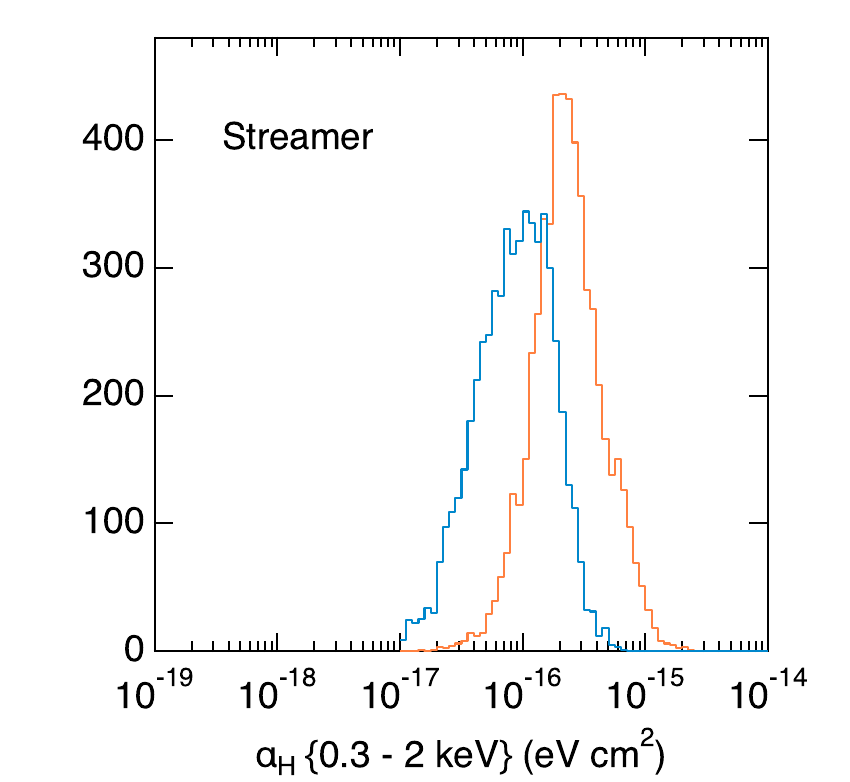}
\noindent\includegraphics[height=3cm, width=0.24\textwidth]{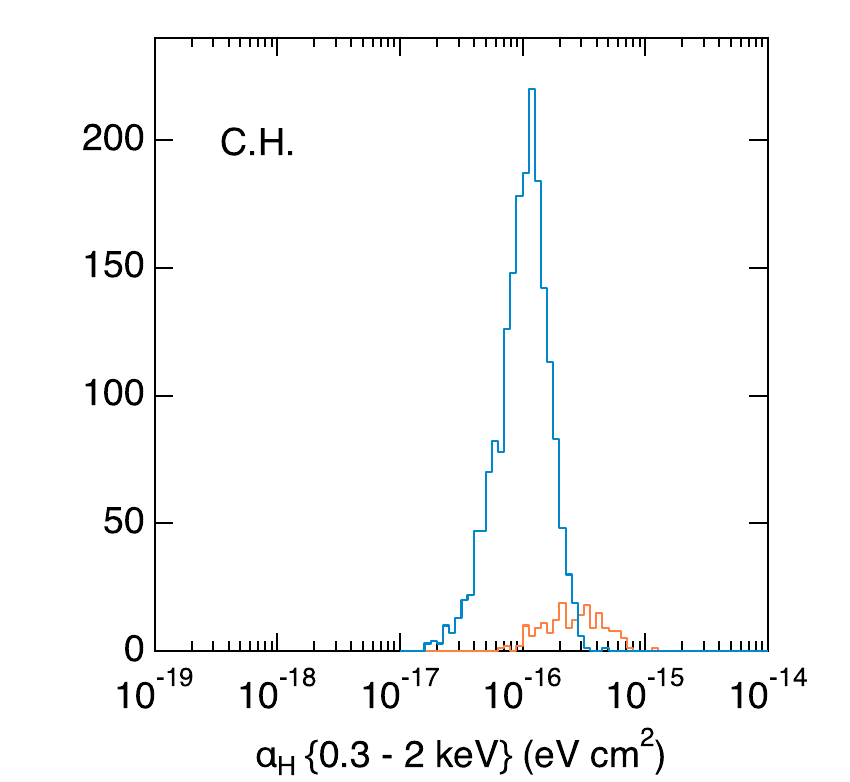}
\noindent\includegraphics[height=3cm, width=0.24\textwidth]{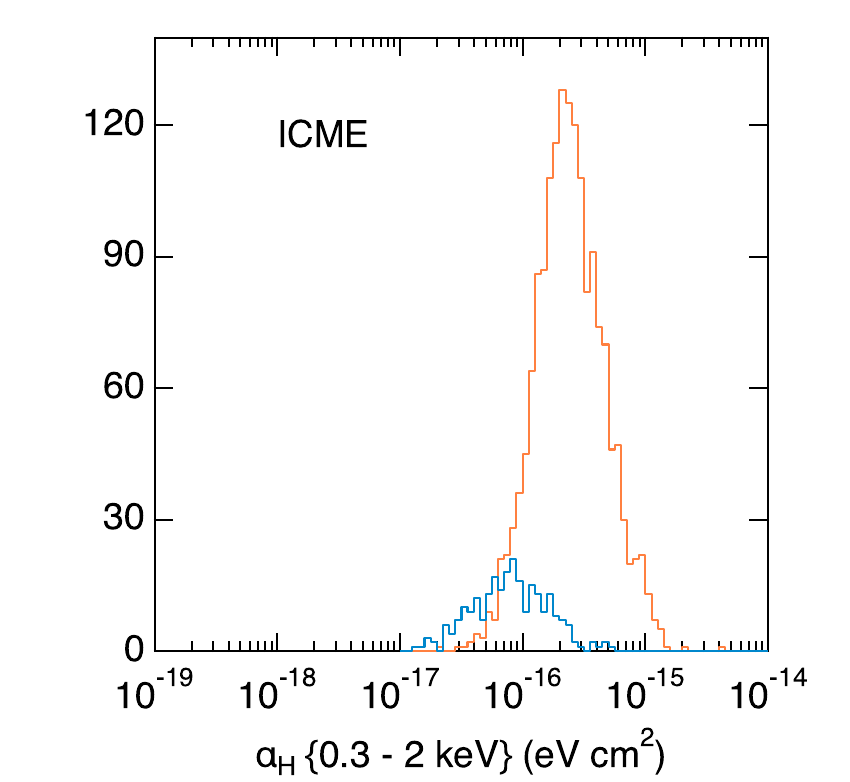}

\noindent\includegraphics[height=3cm, width=0.24\textwidth]{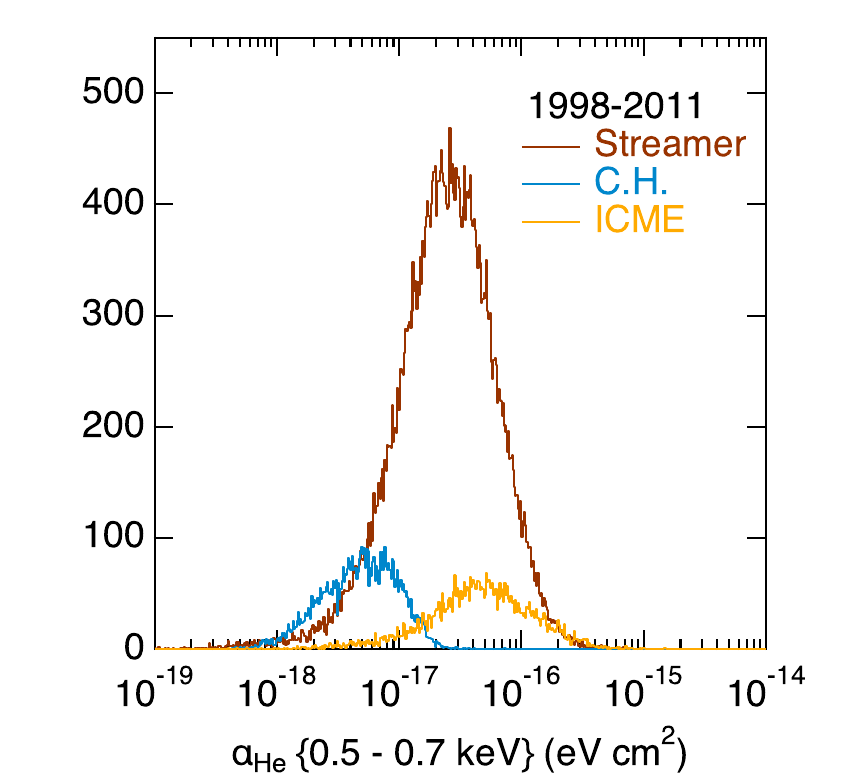}
\noindent\includegraphics[height=3cm, width=0.24\textwidth]{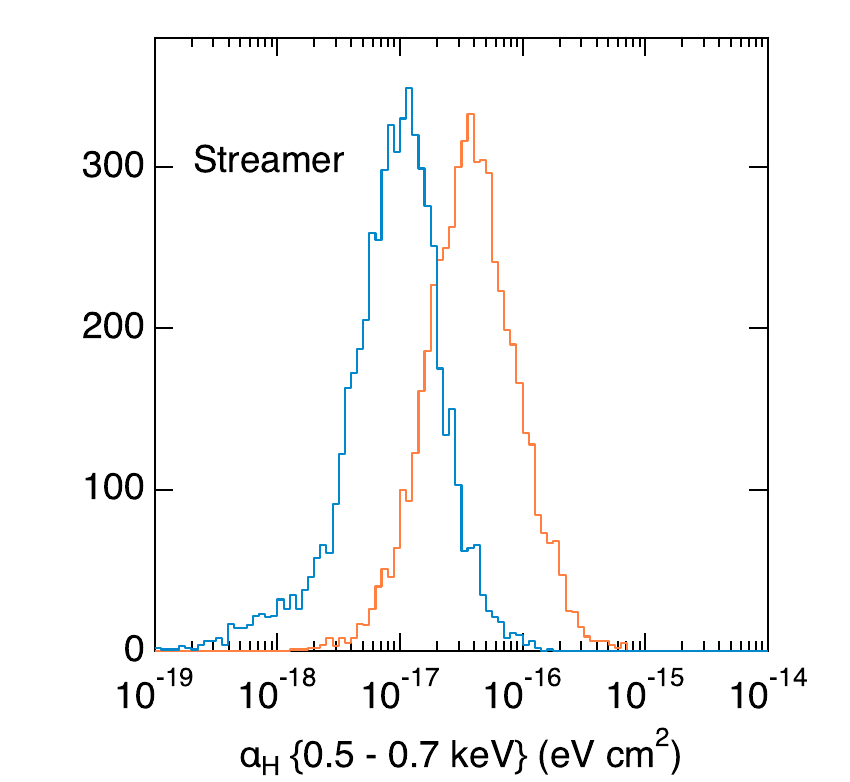}
\noindent\includegraphics[height=3cm, width=0.24\textwidth]{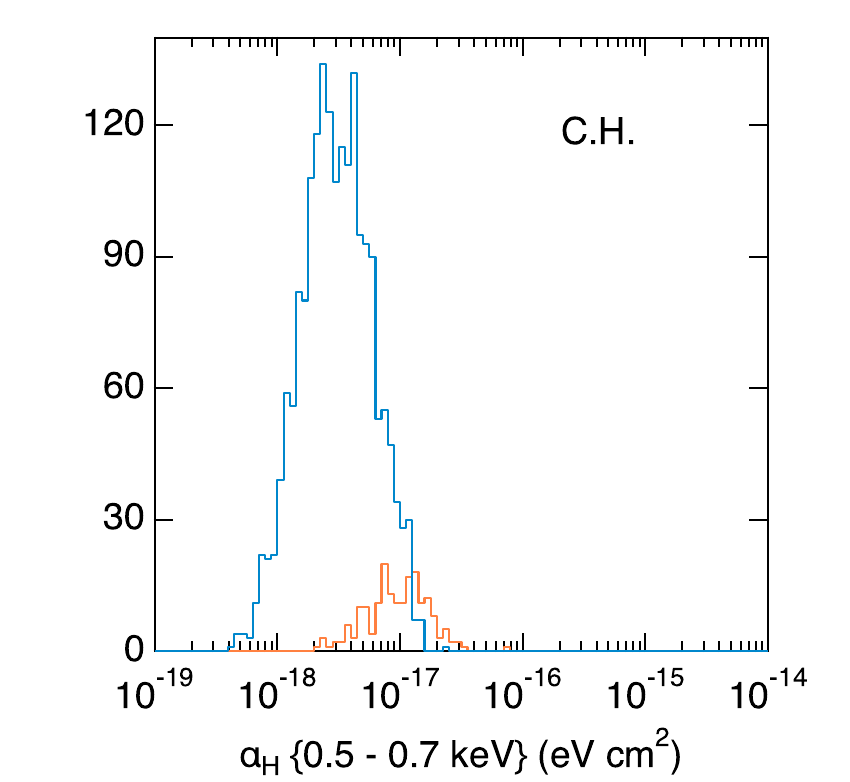}
\noindent\includegraphics[height=3cm, width=0.24\textwidth]{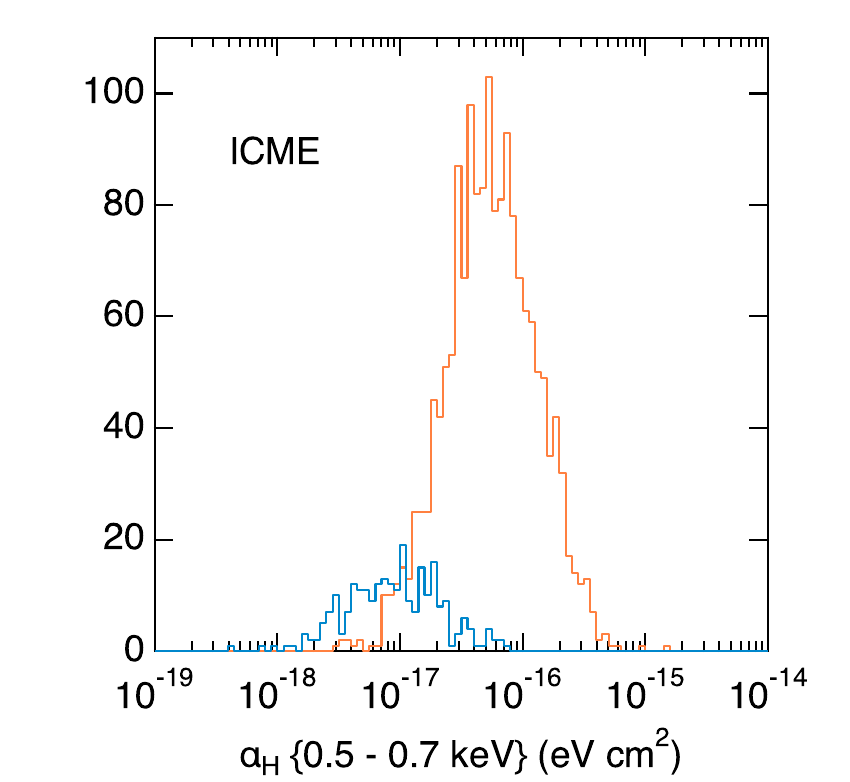}
\caption{From top to bottom, histograms (in number of occurrences) for the $\alpha_H$ 0.1-2 keV, 0.3-2 keV, 0.5-0.7 keV range, and for the $\alpha_{He}$ 0.1-2 keV, 0.3-2 keV, 0.5-0.7 keV range. The first column includes histograms for streamer (dark red), coronal hole (blue) and ICME (yellow) SW for the complete ACE 1.1 database (1998-2011). The second, third and fourth columns include histograms for the streamer, C.H. and ICME types of SW respectively, separated in solar maximum (orange) and solar minimum (blue). The red and blue vertical lines in panel three of the first column correspond to the \protect\citeA{Whittaker2016} values for the same energy range.}
\label{FigAlpHists}
\end{figure}

We calculate the compound cross-section $\alpha$ based on equation \ref{eqAlpha}, where the $[\frac{X^{q+}}{p}]$ ratio is calculated as $[\frac{X^{q+}}{p}] = [\frac{X^{q+}}{O}][\frac{O}{He}][\frac{He}{p}]$, from the ACE SWICS and SWEPAM data presented previously\footnote{The combination of SWICS and SWEPAM data with respect to the He$^{++}$ densities should be treated with caution, as the two instruments may have systematic differences, especially at the lower density regimes (e.g. CH). The SWEPAM n$_{He}$ mode values are provided for comparison. Streamer: 0.135$^{+0.196}_{-0.083}$, CH: 0.081$^{+0.093}_{-0.045}$, ICME: 0.169$^{+0.380}_{-0.119}$.}. Figure~\ref{FigAlphaTS} shows the 27-day average values for the compound cross-sections with H and He in three energy ranges. We have chosen the full pass-band of SXI 0.1-2.0 keV, a slightly narrower band 0.3-2.0 keV that may be applied to the XMM-Newton PN and MOS detectors, and a very narrow band around the oxygen lines 0.5-0.7 keV. In Figure \ref{FigAlpHists} we compare the $\alpha$ distributions (the histograms represent the number of occurrences) per solar wind type and solar activity period for these bands. In Table \ref{TabAlphas} we list the mode values of the distributions presented in Figure \ref{FigAlpHists} for streamer, CH and ICME solar wind type for the complete ACE 1.1 database, as well as for the solar maximum and solar minimum periods. The uncertainties are again determined as the dispersion (width) of the distribution at half maximum.

During the full ACE 1.1 database period, the broad band calculations (0.1 - 2 keV) show little variations with the solar wind type, as it would be expected, since the blend of spectral lines of several ion charge states smooths out any individual variations. The values are of the same order as the \citeA{Cravens2001} empirical value of 6$\times$10$^{-16}$ for photon energies $>$ 0.1 keV. However, if we compare to the $\alpha$ calculated based on the SC00 charge-state abundances and cross-sections (Table \ref{TabAlphas}), our values are more than 2 times lower. This is most probably due to the higher cross-sections used in the SC00 study, since we have demonstrated that the ion abundances are similar in most cases between the SC00 and ACE 1.1 analysis. 

The $\alpha$ differences between solar wind types become more noticeable as the spectral range becomes narrower. The 0.5 - 0.7 keV band that includes mostly the oxygen lines (and some nitrogen) shows the most variability with a factor of ~1.7 increase between streamer-type and ICME-type solar wind, and a factor of 4-6 lower for the CH-type compared to streamer. 

\citeA{Whittaker2016} had calculated $\alpha$ values for the 0.5 - 0.7 keV band based on the ACE O$^{7+}$ and O$^{8+}$ data. They split the solar wind data by velocity (with a cutoff at 500 km/s) and found modal values of 3.1$\times$10$^{-17}$ and 6.1$\times$10$^{-17}$ eV cm$^2$ for slow and fast solar wind respectively (red and blue vertical lines in the third panel of the first column in Figure \ref{FigAlpHists} respectively). These values are inconsistent with the values we find in this analysis, and this may be for several reasons: (i) the velocity cutoff choice by \citeA{Whittaker2016} in contrast with our selection criteria here, (ii) the fact that their calculations do not include the faint nitrogen lines around 0.5 keV as is the case in our analysis, and (iii) their use of cross-sections/line emission probabilities from \citeA{Bodewits2007}, which may be somewhat different from the ones considered here. However, we estimate that the most important reason for the difference in their values compared to ours is the use of linear bins/gaussian fits for their histograms of the $\alpha$ distributions instead of the logarithmic bins/log-normal distributions we employ here. Most puzzling is the fact that their fast (CH) solar wind $\alpha$ is higher than their slow (streamer) solar wind value, although they use oxygen charge-state abundances that are systematically less abundant in the fast solar wind. Moreover, in their table 1, the $\alpha$ value as a function of velocity shows a maximum of 8.2$\times$10$^{-17}$ eV cm$^2$ at 400 km/s. It is not clear to us what are the reasons for these inconsistencies.

\begin{sidewaystable}
\caption{Compound cross-sections of SWCX with H and He, in units of 10$^{-16}$ eV cm$^2$, for streamer (Slow), CH (fast) and ICME solar wind for different solar activity periods. The uncertainties represent the dispersion (width) of the distribution at half maximum.}
 \centering
 \begin{tabular}{l l c c c c c c c c c}
 \hline
&Energy   &\multicolumn{3}{c}{Full period (1998-2011)}&\multicolumn{3}{c}{Max (2000-2002)}& \multicolumn{3}{c}{Min (2008-2010)}\\
&Range (keV) &Str. &CH &ICME &Str. &CH &ICME &Str. &CH &ICME \\
 \hline
&0.1 - 2.0 & 7.8 $^{+  7.0}_{-  3.7}$ & 7.1$^{+  4.7}_{-  2.9}$ & 8.8$^{+  8.8}_{-  4.4}$ & 8.5$^{+  8.2}_{-  4.3}$& 9.2$^{+  9.4}_{-  4.6}$& 8.6$^{+  9.0}_{-  4.5}$& 3.7$^{+  4.6}_{-  2.1}$& 4.7$^{+  3.2}_{-  1.7}$& 3.2$^{+  4.2}_{-  1.8}$\\
$\alpha_H$ &0.3 - 2.0 & 4.5$^{+  4.3}_{-  2.2}$ & 3.2$^{+  2.3}_{-  1.3}$ & 5.6$^{+  6.2}_{-  3.0}$ & 5.0$^{+  4.9}_{-  2.6}$& 4.6$^{+  5.3}_{-  2.4}$& 5.6$^{+  6.2}_{-  3.0}$& 1.9$^{+  2.8}_{-  1.1}$& 2.0$^{+  1.3}_{-  0.8}$& 1.7$^{+  2.3}_{-  0.9}$\\
&0.5 - 0.7 & 1.2$^{+  1.9}_{-  7.5}$ & 0.3$^{+  0.3}_{-  0.1}$ & 2.1$^{+  3.4}_{-  1.3}$ & 1.7$^{+  2.3}_{-  1.0}$& 0.5$^{+  0.4}_{-  0.2}$& 2.3$^{+  3.2}_{-  1.4}$& 0.5$^{+  0.6}_{-  0.3}$& 0.2$^{+  0.2}_{-  0.1}$& 0.4$^{+  0.8}_{-  0.3}$\\
\hline
&0.1 - 2.0 & 4.6 $^{+  4.2}_{-  2.3}$ & 4.8$^{+  3.0}_{-  1.9}$ & 4.7$^{+  5.2}_{-  2.5}$ & 4.9$^{+  5.0}_{-  2.4}$& 6.2$^{+  5.5}_{-  2.9}$& 4.5$^{+  4.8}_{-  2.3}$& 2.3$^{+  2.9}_{-  1.3}$& 3.2$^{+  2.1}_{-  1.2}$& 2.0$^{+  2.6}_{-  1.2}$\\
$\alpha_{He}$ &0.3 - 2.0 & 1.9$^{+  2.0}_{-  1.0}$ & 1.7$^{+  1.3}_{-  0.7}$ & 2.2$^{+  2.4}_{-  1.2}$ & 2.1$^{+  2.3}_{-  1.1}$& 2.5$^{+  2.7}_{-  1.3}$& 2.2$^{+  2.4}_{-  1.1}$& 0.9$^{+  1.2}_{-  0.5}$& 1.0$^{+  0.7}_{-  0.4}$& 0.8$^{+  1.0}_{-  0.4}$\\
&0.5 - 0.7 & 0.3$^{+  0.4}_{-  0.2}$ & 0.05$^{+  0.07}_{-  0.03}$ & 0.5 $^{+  0.9}_{-  0.3}$ & 0.4$^{+  0.6}_{-  0.2}$& 0.09$^{+  0.09}_{-  0.04}$& 0.5$^{+  0.9}_{-  0.3}$& 0.09$^{+  0.10}_{-  0.06}$& 0.03$^{+  0.04}_{-  0.02}$& 0.09$^{+  0.16}_{-  0.06}$\\
\hline
&0.1 - 2.0 & 22.8 & 12.2 & & & & & & & \\
$\alpha_{SC}^{(a)}$ &0.3 - 2.0 & 16.2 & 6.8 &  & & & & & & \\
&0.5 - 0.7 & 8.9 & 1.3 &  & & & & & & \\
\hline
 \multicolumn{11}{l}{$^{(a)}$Calculated based on the $[\frac{X^{q+}}{O}]$ and cross-sections values from Table 1 in SC00.}\\
\end{tabular}
\label{TabAlphas}
 \end{sidewaystable}

The solar minimum values per solar wind type are systematically lower compared to solar maximum, which is in agreement with the depleted abundances found in the ACE data analysis in Section \ref{SecSW}. The streamer population has a compound cross-section from 2 to 4.5 times higher in solar maximum compared to solar minimum depending on the energy band and neutral target. The CH population shows changes of a factor of 2 to 3, and the ICME population shows changes of a factor of 2.3 to 5.8. 

It is also worth noting that in some occasions, in particular for the 0.1-2 keV band in solar maximum and solar minimum, the CH populations exhibit higher compound cross-sections compared to the streamer and ICME populations (Table \ref{TabAlphas}). This is probably due to the fact that these bands are populated by spectral lines produced from source ions with lower charge-states (e.g. C$^{5+}$, O$^{6+}$) that are more abundant in the CH-type solar wind, as shown in Table \ref{Tab2}. The decrease of higher q ions or increase in lower q ions for the CH wind, compared to streamer and ICME winds is stronger for lower Z elements and less apparent for higher Z elements. However, since the lower Z elements are generally more abundant than higher Z elements, and preferentially populate the lower energy portion of the spectrum, small effects in the ion ratios for the lower Z elements are seen more easily in the $\alpha$ values.

\section{Conclusion}\label{SecConclusion}
We have compiled ion charge state abundances from the ACE 1.1 database between 1998 and 2011, charge-exchange cross-sections from the KRONOS database, and line emission probabilities from literature. We have calculated the compound cross-sections $\alpha$ for charge-exchange collisions with H and He atoms for broad (0.1 - 2 keV), average (0.3 - 2 keV) and narrow (0.5 - 0.7 keV) spectral ranges, for streamer, CH and ICME solar wind types and in different solar cycle periods.

We find that for broad band ranges (0.1 - 2 keV), there is little variation in $\alpha$ for the different solar wind types in each solar period. The distinction between solar wind types becomes significant as we narrow down the spectral band to single ion emission range, such as the oxygen band (0.5 - 0.7 keV). This is consistent with previous studies showing that the SWCX signal is more influenced by individual ion variations when analysing bands where only few ions dominate the spectrum, as opposed to broad bands where blends of many ion spectral lines are measured \cite{Kuntz2015}. Most notably, we find variations between solar maximum and solar minimum periods for each solar wind type separately, due to sharp changes in ion abundances measured by ACE/SWICS, as demonstrated in this and previous studies \cite{Lepri2013}.

Using the compound cross-sections in broad X-ray bands is a convenient method to quickly link the SWCX signal in magnetospheric calculations with global increases of solar wind proton flux. However, it has been demonstrated in previous studies that the individual ion abundances may impact the SWCX signal in a more significant way \cite<e.g.,>[]{Zhang2022}. Therefore it is important to combine several approaches allowing for quick analyses of the magnetosphere's dynamic response (e.g. with MHD models), and more detailed modeling focusing on individual ions and the way they evolve around the magnetospheric boundaries, or precipitate through the cusps (e.g. with test-particle models). The SMILE/SXI instrument has both a broad energy range (0.1 - 2 keV) and enough spectroscopic resolution to separate several strong lines (e.g. from O, C, N and Ne ions), and will certainly benefit from these combined approaches.

In that context, and in view of future X-ray missions (e.g., LEM, ATHENA) that will provide high-resolution spectra of the SWCX emission in the 0.1 - 2 keV range, it is crucial to improve developments in two areas. First, we need to improve the databases of cross-sections and emission probabilities for complex ion structures, such as the ions Mg, Si, S, Fe, populating the low-energy band 0.1 - 0.3 keV. For this, we need accurate quantum calculations, corroborated by experimental measurements in collision energies similar to astrophysical conditions. In addition, continuous monitoring of the solar wind ion composition is a key link between the solar wind effects through interplanetary space and in approach of the magnetosphere to the SWCX signal measured near the magnetospheric boundaries and the cusps. With the declining performances of ACE/SWICS since 2011, it becomes imperative that new missions allowing for monitoring the solar wind ion composition are developed and, if possible, launched in tandem with X-ray observatories.

\acknowledgments
This work was supported by CNES. I wish to thank the SWEPAM and SWICS instrument teams for providing the ACE data through the ACE Science Center: http://www.srl.caltech.edu/ACE/ASC/index.html~. I am grateful to Sue Lepri, Jim Raines, Ruedi von Steiger and Liang Zhao for invaluable discussions on the SWICS composition data.
I am very thankful to Renata Cumbee for her advice on the use of the cross sections in the KRONOS charge transfer database. I cannot omit to thank Vasili Kharchenko for always happily answering my endless questions on line emission probabilities. Finally, I deeply appreciate the two referees' thorough evaluations that greatly improved this paper.
 \\



%
 \bibliography{CXAlpha_SMILE}
%




%
%
%
%
%

\end{document}